\documentclass[iop]{emulateapj}

\usepackage{natbib}
\usepackage{graphicx}
\bibpunct{(}{)}{;}{a}{}{,}

\begin{document}
%
\title{Relationship between Hubble type and spectroscopic class
in local galaxies}
\author{  
	J.~S\'anchez~Almeida\altaffilmark{1,2},
	J.~A.~L.~Aguerri\altaffilmark{1,2},  
	C.~Mu\~noz-Tu\~n\'on\altaffilmark{1,2}, 
	and 
	M.~Huertas-Company\altaffilmark{3,4}
          }
\altaffiltext{1}{Instituto de Astrof\'\i sica de Canarias, E-38205 La Laguna,
Tenerife, Spain}
\altaffiltext{2}{Departamento de Astrof\'\i sica, Universidad de La Laguna,
Tenerife, Spain}
\altaffiltext{3}{GEPI, Paris-Meudon Observatory 5, Place Jules Janssen, 92190
Meudon, Paris, France}
\altaffiltext{4}{Universit\'e Paris Diderot, 75205 Paris Cedex 13, France}
\email{jos@iac.es, jalfonso@iac.es, cmt@iac.es, marc.huertas@obspm.fr}
\begin{abstract}
We compare the Hubble type and the spectroscopic class of the galaxies with spectra in 
SDSS/DR7.  As it is long known, elliptical galaxies tend to be red whereas spiral galaxies tend
to be blue, however,  this relationship presents a large scatter,  
which we measure and quantify in detail for the first time.
We compare the Automatic Spectroscopic K-means based classification
(ASK) with most of the commonly used morphological classifications. 
Despite the degree of subjectivity involved in  the morphological classifications, all of them provide 
consistent results. 
Given a spectral class, the morphological type wavers with
a standard deviation  between 2 and 3 $T$~types, 
and the same large dispersion characterizes the variability of spectral classes fixed the 
morphological type.
The distributions of Hubble types given an ASK class are very skewed --
they present long tails that go to the late morphological 
types for the red galaxies, and to the early morphological
types for the blue spectroscopic classes.
The scatter is not produced by problems in the classification, and it 
remains
when particular subsets are considered -- low and high galaxy masses, 
low and high density environments, barred and non-barred galaxies,
edge-on galaxies, small and large galaxies, or when a volume limited
sample is considered. 
A considerable fraction of the red galaxies are spirals 
(40--60\%), but they never present very late Hubble types 
(Sd or later).
Even though red spectra are not associated
with ellipticals, most ellipticals do have red
spectra: 97\% of the ellipticals in the morphological
catalog by \citeauthor{nai10},
used here for reference, belong to ASK~0, 2 or 3. It contains
only a 3\% of  blue ellipticals.
The galaxies in the green valley class (ASK~5) are mostly spirals,
and the AGN class (ASK~6) presents a large scatter of Hubble
types from E to Sd.
We investigate variations with redshift using a
volume limited subsample mainly formed by luminous 
red galaxies. From redshift $0.25$ to now the 
galaxies redden from ASK 2 to ASK 0, as expected from 
the passive evolution of their stellar populations.
Two of the ASK classes (1 and 4) gather edge-on spirals, and they
may be useful in studies requiring knowing the intrinsic shape of 
a galaxy (e.g., weak lensing calibration).
\end{abstract}


\shorttitle{Hubble type vs ASK spectral class}
\shortauthors{S\'anchez Almeida et al.}

%


\section{Introduction}\label{introduction}
It is long known how the Hubble sequence correlates
with the colors of the galaxies. Ellipticals tend
to be red whereas spirals tend to be blue 
\citep[e.g.,][]{hub36, hum31,mor57,aar78,ber95,zar95}.
It is also well known that the relationship
is far from being one-to-one \citep[e.g.][]{con95,sod97,for04,fer06}. 
There are red spiral galaxies 
\citep[e.g.,][]{van76,dre99,den09,mas10}, 
as well as blue ellipticals \citep[e.g.,][]{den09,sch09b,kan09,hue10a}. 
The correlation changes with time, and 
some morphologies now rare were
not unusual early on \citep[e.g.,][]{res03,elm05}. 
Blue galaxies become more common as the universe gets 
younger \citep[e.g.,][]{lil95,mad96,del10}, 
even though many ellipticals and spirals were already in place 
at high redshift 
\citep[e.g.,][]{van02,agu02,tru04}.
The relationship gets fuzzier with increasing lookback time, 
to perhaps disappear in the early universe  
\citep{con06,hue09}.

In principle, the origin of the relationship between 
spectrum and morphology is relatively well understood.
Ellipticals formed their stars long ago, therefore, their
stellar populations are aged and so red. Spirals are still
forming stars, and massive young stars that are blue
contribute significantly to the integrated galaxy spectrum.
The details of the relationship are not so clear, though. 
It remains unknown why local galaxies are 
grouped into two main colors \citep[e.g.,][]{str01,bla03,bla09} while 
their shapes follow a continuous distribution of Hubble types. 
The reasons for having outliers from the main relationship
are even more unclear.  
Do outliers represent unusual galaxies living outside the main stream,
or are they transients during the regular galaxy evolution?
Galaxies evolve in spectrum and in shape during their lifetimes, but we 
ignore whether the morphological changes lead or trail the spectral changes.

The advent of new data sets allows us characterizing the
relationship between Hubble type and spectroscopic
class with unprecedented detail. This is the purpose of our
work. We compare the spectroscopic classification by \citet[][]{san10}
with various morphological catalogs existing in 
the literature \citep[][]{dev91,ken92,fuk07,nai10,lin10,hue11}.
The  Automatic Spectroscopic K-mean-based  classification
by \citeauthor{san10}, ASK, comprises the $\sim 10^6$ galaxies 
with spectra in Sloan Digital Sky Survey Data Release seven 
\citep[SDSS/DR7,][]{sto02,aba09}.
We use ASK because it provides a comprehensive description 
of all the spectral types existing in nearby galaxies. 
It is finer than other alternatives involving large data sets 
\citep[e.g., principal component analysis,][]{yip04}, and the
option of using color cuts and line ratios would require
additional work to assess the completeness of the approach.
In the vein of all previous studies comparing spectroscopic 
and morphological classifications,
we find a general trend with large scatter. 
The work describes and quantifies
such relationship, as a constraint to be used elsewhere 
to test galaxy evolution theories. 

The paper is organized as follows: \S~\ref{sec_ask}
introduces the spectral classification
to make the paper comprehensive. The empirical comparison
between spectral classes (spectro-class) 
and Hubble types (morpho-types) is carried out in 
\S~\ref{sec_ask_vs_hubble}. We use the 
catalog by \citet{nai10} as reference since 
it provides an optimal compromise between
enough galaxies and detailed morphological types
(\S~\ref{sect_nair_abra}),
however, many of the morphological catalogs 
commonly used are analyzed as well
(\S~\ref{morph_fuk} \citeauthor{fuk07}~\citeyear{fuk07};
\S~\ref{sec_rc3} \citeauthor{dev91}~\citeyear{dev91};
\S~\ref{svm_sect} \citeauthor{hue11}~\citeyear{hue11};
\S~\ref{sect_gzoo} \citeauthor{lin10}~\citeyear{lin10};
and 
\S~\ref{sec_ken} \citeauthor{ken92}~\citeyear{ken92}). 
The morphological classifications 
have a large degree of subjectivity. The hazy logic
behind the human pattern recognition skills is responsible for
the types, either directly, or through training sets in machine learning 
automatic procedures. In order to be more quantitative, we try to compare the 
ASK classes with morphologically related parameters that can be measured directly 
from galaxy images (\S~\ref{quantify}), and which have been used
in the so-called quantitative morphology 
\citep[see, e.g.,][]{ode02,cal04,con06}. 
Possible variations with redshift of the color-shape relationship are briefly considered
in \S~\ref{sect_redshift}.
The results are analyzed and summarized in \S~\ref{discussion}.
Distances and look-back times 
are based on 
a Hubble constant $H_0= 70$~km\,sec$^{-1}$\,Mpc$^{-1}$.
All our galaxies have redshifts~$\le 0.25$.

%

%
%
\section{The ASK spectral classification}\label{sec_ask}

The ASK spectral classification is detailed in 
\citet{san10}.  For the sake comprehensiveness, however, this section 
summarizes its main properties. 
ASK considers all the galaxies with spectra 
in SDSS/DR7. Those with redshift smaller than 
0.25 are transformed to a common rest-frame 
wavelength scale, and then re-normalized to
the integrated flux in the SDSS $g$-filter. 
These two are the only manipulations the
spectra undergo before classification. 
We wanted the classification to be driven only 
by the shape of the visible spectrum
(from 400 to 770 nm), and these two 
corrections remove obvious undesired dependencies 
of the observed spectra on redshift and galaxy apparent
magnitude. We deliberately avoid correcting for other 
effects requiring modeling and assumptions 
(e.g., dust extinction, seeing, or aperture effects). This approach 
is in the spirit of the rules for a good classification put forward 
by \citet{san05}, where he points out that physics must not drive a 
classification. Otherwise the arguments become circular when the 
classification is used to drive physics.
Obviously this approach does
not imply disregarding the fairly complete
understanding of galaxy physics we have today. It 
just separates the interpretation of the spectra from 
determining their observed shapes.
The employed classification algorithm, k-means, is a robust workhorse
that makes it doable the simultaneous classification of the
full data set. It is commonly employed in data mining, machine learning, 
and artificial intelligence \citep[e.g.,][]{eve95,bis06}, 
and it guarantees that similar rest-frame spectra belong to 
the same class.
99\% of the galaxies can be  assigned to only 17 major 
classes, with 11 additional minor classes including the
remaining 1\%.
It is unclear whether the ASK classes represent
genuine clusters in the 1637-dimensional classification
space, or if they partake a continuous distribution -- probably 
the two kinds of classes are present
\citep[see][]{san10,asc11}.
All the galaxies in a class have very similar
spectra, which are also similar to the class template spectrum
formed as the average of all the spectra of the galaxies 
in the class. These template spectra vary smoothly and 
continuously. They were labeled according to the $u-g$ color, 
from the reddest, ASK~0, to the bluest, 
ASK~27. The use of numbers to label the classes does not
implicitly assumes the spectra to follow
a one dimensional family. The numbers only
name the classes. The sorting (and, so, the naming) 
would have been slightly different using other bandpasses
to define colors. In general, however, the smaller the ASK class
the redder the full spectrum, and we often use the 
terms red and blue referring to low and high
ASK numbers, respectively. The ASK classification 
of all the galaxies with spectra in SDSS/DR7 is publicly 
available\footnote{
{\tt ftp://ask:galaxy@ftp.iac.es/}\\
{\tt http://sdc.cab.inta-csic.es/ask/index.jsp} in the 
Spanish Virtual Observatory,
and also in SDSS CasJobs as 
{\tt public.jalmeida.ask} and {\tt public.jalmeida.highz\_ask}.
}.
In this paper we only employ the $\sim 7\cdot 10^5$ SDSS/DR7 galaxies having
redshifts~$\le 0.25$.
%
\section{Morphological type versus spectroscopic class}\label{sec_ask_vs_hubble}

Eyeball morphological classifications are based on 
factors like the presence or not of a disk, the bulge-to-disk ratio, 
the presence of arms, their organization, and so on.  Obviously,
they entail a significant degree of subjectivity. 
In order to evaluate the effect of this source of error on the 
relationship between morphological type and spectroscopic class, 
we use a number of different classifications. They are not 
homogeneous, having diverse finesse and, therefore, based on 
slightly different criteria --  e.g., \citet[][]{nai10} employ 14 types whereas \citet{lin10} divide 
all galaxies into only two bins, ellipticals and spirals. 
Following Table~1 in \citet{nai10},  
our Table~\ref{equiv_morpho} provides a rough equivalence 
between the different morphological classifications employed here.
The classical Third Reference Catalog by \citet[][RC3]{dev91}
is used as guide in Table~\ref{equiv_morpho}. The criteria 
that define the types in this  catalog are summarized in 
\S~\ref{sec_rc3}, where we employ RC3 to provide the morphological types. 
\begin{deluxetable*}{cccccccccccccccccc}
\tabletypesize{\scriptsize}
\tablecolumns{19}
\tablewidth{0pc}
\tablecaption{Equivalence between the different morphological 
classification schemes used in the work}
\tablehead{
\colhead{}&
\colhead{c0}&	\colhead{E0}&	\colhead{E+}&	\colhead{S0-}&	\colhead{S0}&
\colhead{S0+}&	\colhead{S0/a}&	\colhead{Sa}&	\colhead{Sab}&	\colhead{Sb}&
\colhead{Sbc}&	\colhead{Sc}&	\colhead{Scd}&	\colhead{Sd}&	\colhead{Sdm}&
\colhead{Sm}&	\colhead{Im}	
}
\startdata
\citet{dev91}\tablenotemark{a}&	-6&	-5&	-4&	-3&	-2&	-1&
0&	1&	2&
3&	4&	5&	6&	7&	8&	9&	10\\	
%
\citet{fuk07}&	E&	E&	E&	S0&	S0&	S0&	S0&	Sa&
Sa&	Sb&	Sb&	Sc&	Sc&	Sd&	Sd&	Im&	Im\\	
\citet{nai10}&	E&	E&	E&	ES0&	S0&	S0&	S0a&
Sa&	Sab&	Sb&	Sbc&	Sc&	Scd&	Sd&	Sdm&	Sm&	Im\\
\citet{hue11}&E&E&E&E&S0&S0&S0&Sab&Sab&Sab&Scd&Scd&Scd&Scd&Scd&Scd&Scd\\
\citet{lin10}\tablenotemark{b}&E&E&E&Sp&Sp&Sp&Sp&Sp&Sp&Sp&Sp&Sp&Sp&Sp&Sp&Sp&Sp
%
\enddata
\tablecomments{The table has been adapted 
        from \citet{nai10}, Table~1. The classification in the
        first row is not used here, but it remains as in the original table for  
        cross-reference.
        }
\tablenotetext{a}{RC3}
\tablenotetext{b}{Galaxy Zoo 1}
\tablenotetext{------}{{\bf To be printed in portrait mode}}
\label{equiv_morpho}
\end{deluxetable*}

%
\subsection{Hubble types derived by \citet{nai10}}\label{sect_nair_abra}
The morphological classification by \citet{nai10} serves as reference. Although 
all the analyzed classifications provide consistent results, the one by 
\citeauthor{nai10} is selected for in-depth
study because it represents the best compromise 
between volume and finesse. 
It contains 14034 galaxies, distinguishes between
14 morphological types (see Table~\ref{equiv_morpho}),
and contains additional information like the 
presence of bars, or the galaxy environment.
In addition, the classification by  \citeauthor{nai10} is almost
magnitude limited, therefore, it can be employed to 
model the results corresponding to
a volume limited sample. The classification
considers all the galaxies in SDSS/DR4 with  $g < 16$ and redshift 
$\leq 0.1$. The redshift upper limit removes almost no galaxy from 
the original pool therefore, in practice, the sample is 
apparent magnitude limited.  

\begin{figure}
        \includegraphics[width=.5\textwidth]{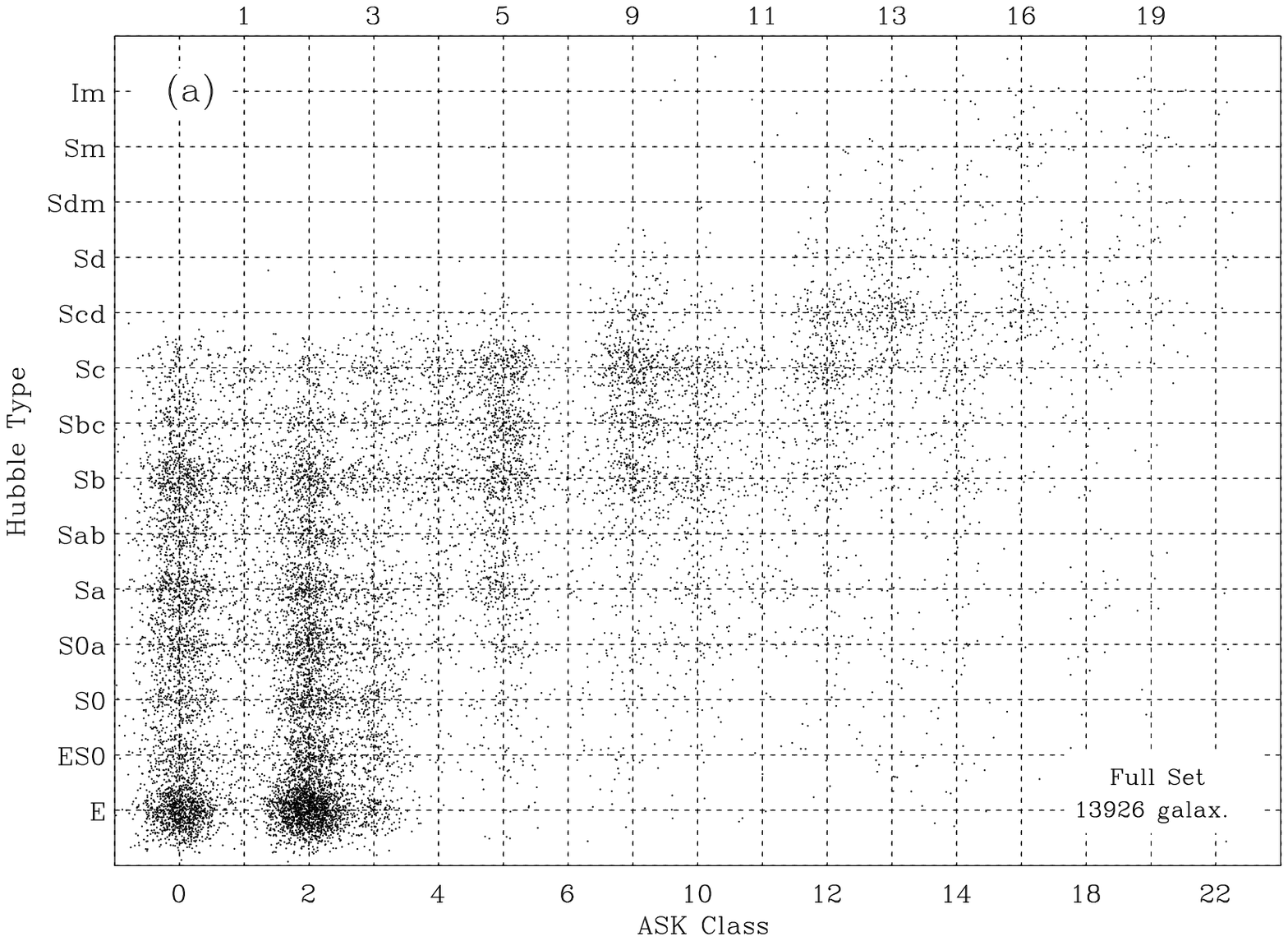}\\
        \includegraphics[width=.5\textwidth]{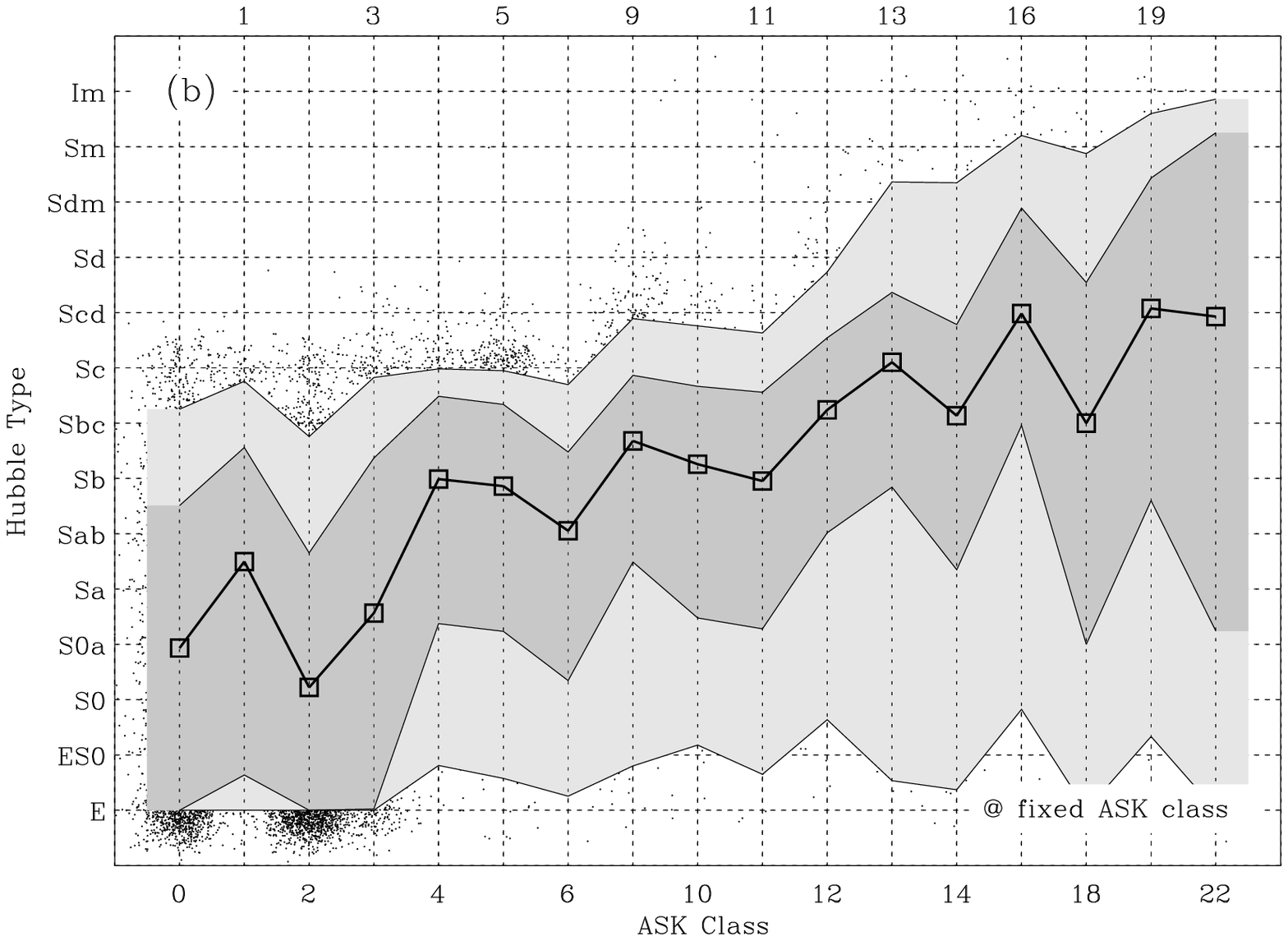}\\
        \includegraphics[width=.5\textwidth]{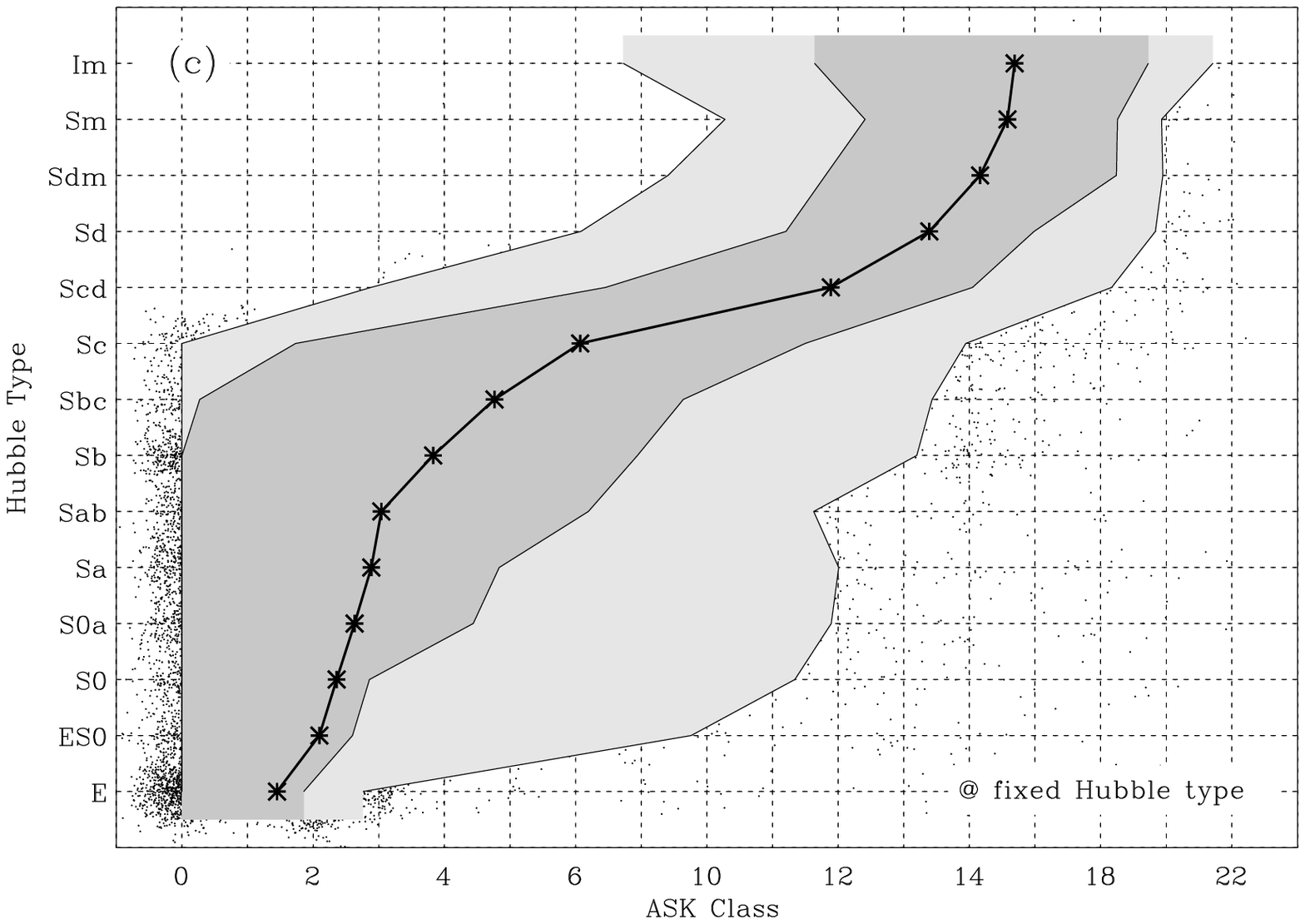}\\
        \caption{
        Hubble type vs ASK class for the full set of 
        galaxies in \citet[][]{nai10}.
        (a) Scatter plot with small noise added to avoid overlapping.
        (b) Average Hubble type given an ASK class (the square symbols), 
        and regions containing 68\% of the galaxies (the dark shaded area) 
        and 95\% of the galaxies (the light shaded area).
        (c) Average ASK class given a Hubble type (the asterisks),
        and regions containing 68\% and 95\% of the galaxies. Abscisae
        include only major ASK classes. 
        }
        \label{nair1}
        \end{figure}

\begin{figure*}
\centering
\includegraphics[width=1.\textwidth,angle=0]{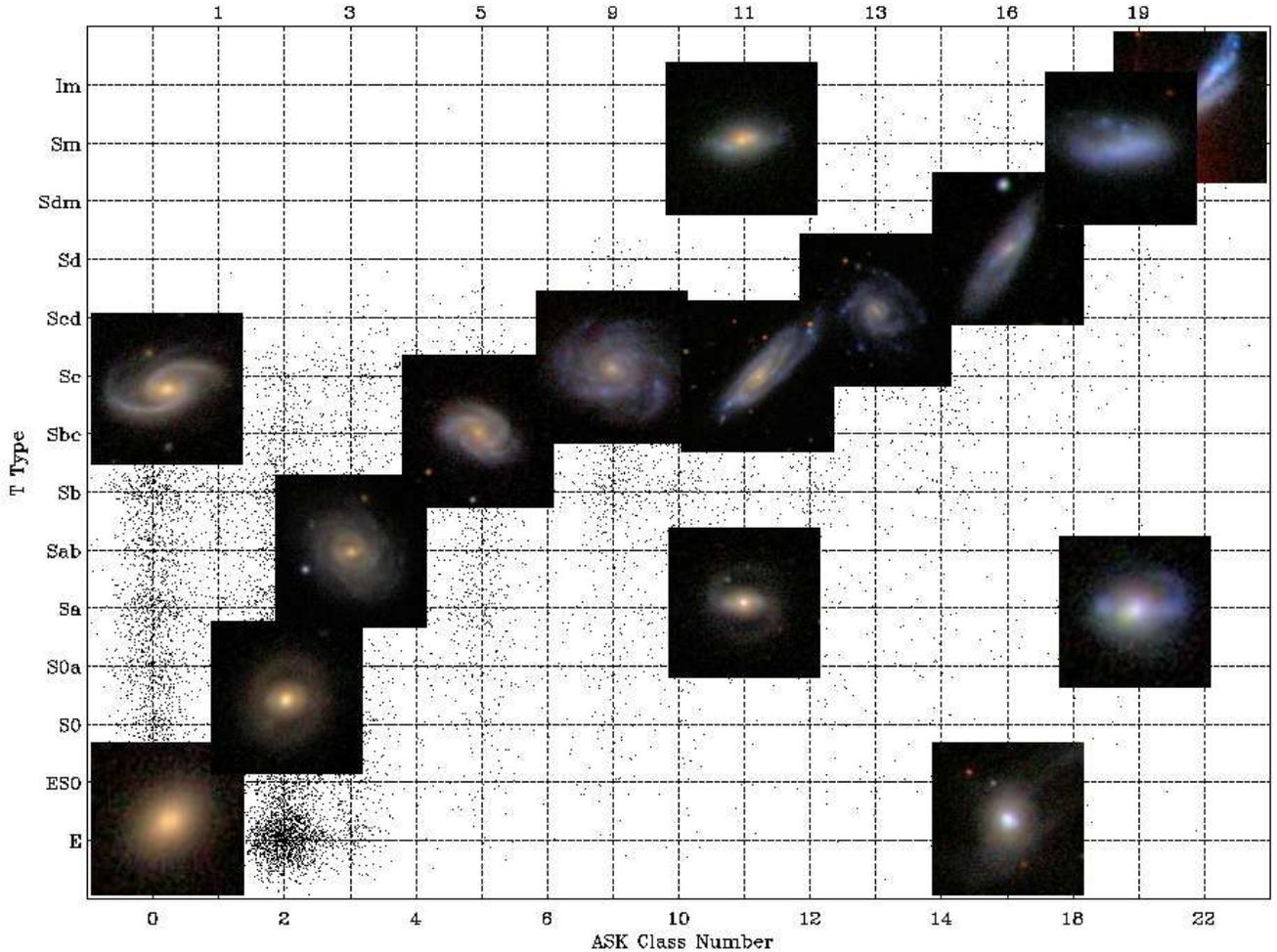}
\caption{
Scatter plot of Hubble type versus ASK class for the
galaxies classified by \citet{nai10} (same as Fig.~\ref{nair1}a).
In addition, we overplot SDSS images of representative galaxies located at 
various places on the diagram. Note the trend from red
early types, to blue late types, roughly following the
diagonal of the diagram. On top of it, however, 
there is a large scatter produced by red spirals 
(e.g., Sc \& ASK~0), and blue ellipticals
(e.g., E \& ASK~16). Only the upper left corner
of the diagram is truly devoid of targets, which would 
correspond to (non-existing) red Sd, Sdm, Im and Sm galaxies.
}
\label{nair4}
\end{figure*}

Figure~\ref{nair1}a shows the scatter plot of Hubble type vs ASK class
for all the galaxies in \citet{nai10}.
As we will do throughout the work, we have added some random 
noise (normally distributed, 0.3 classes standard deviation) 
to the actual positions 
of the galaxies.
The sampling in Hubble type and ASK class is discrete and
so coarse that, unless noise is added, the galaxies overlap not showing up
in scatter plots\footnote{This trick is employed for purely 
aesthetic reasons, and noise is not included when evaluating
the parameters given in the paper.}.
The first obvious result drawn from inspecting
Fig.~\ref{nair1}a is the dispersion of ASK classes
fixed  the morphological type, and vice versa. Such 
dispersion exceeds the uncertainty in Hubble type 
\citep[$\le 0.5$ types --][]{nai10} 
and ASK class \citep[$< 1$ class --][]{san10}.
In order to quantify the dispersion, we make use of the 
two dimensional histogram $N(A,T)$, with $A$ the ASK class, 
and $T$ the Hubble type.
We characterize the properties of the observed distribution
using the first moments of the two marginal distribution 
functions that result from integrating over one of the
two variables. Namely, given a class $A$, 
we calculate the mean $T$~type, $\mu_T(A)$, 
\begin{equation}
\mu_T(A)=N^{-1}_T(A)\,\sum_iT_i\,N(A,T_i),
\end{equation} 
the standard deviation, $\sigma_T(A)$,
\begin{equation}
\sigma^2_T(A)=N^{-1}_T(A)\,\sum_i\big[T_i-\mu_T(A)\big]^2\,N(A,T_i),
\end{equation} 
the skewness\footnote{\label{def_kur}The skewness parameterizes whether 
the distribution is symmetrical about its maximum.
Positive skewness indicates a longer tail to
the right of the distribution maximum, whereas negative skewness 
indicates  a tail extending to the left of the distribution maximum.
The kurtosis parameterizes whether the distribution is more peaked
than a gaussian ($\kappa > 0$) or if it has extended tails 
($\kappa < 0$).},
$\beta_T(A)$,
\begin{equation}
\beta_T(A)=N^{-1}_T(A)\,\sum_i\big\{\big[T_i-\mu_T(A)\big]/\sigma_T(A)\big\}^3\,
N(A,T_i),
\end{equation}
and the kurtosis$^{\ref{def_kur}}$, $\kappa_T(A)$,
\begin{equation}
\kappa_T(A)=N^{-1}_T(A)\,\sum_i\big\{\big[T_i-\mu_T(A)\big]/\sigma_T(A)\big\}^4\,
N(A,T_i)-3.
\end{equation}
The sums are over all $T$~types given an $A$ class,
and the symbol
$N_T(A)=\sum_i\,N(A,T_i)$ stands for 
the total number of galaxies belonging to class $A$.
We also compute the statistical parameters  
for the dispersion among the ASK classes
given a $T$~type. They are formally identical 
to the previous definitions but exchanging $A$ with $T$, i.e.,
$\mu_A(T), \sigma_A(T), \beta_A(T)$ and $\kappa_A(T)$.
Tables~\ref{table_ask} and \ref{table_hubble} list all these
statistical parameters plus the median (type that 
splits the sample into two halves) and the mode (most probable
value). 
Note than one can approximately retrieve the full
distribution from these few moments \citep[e.g.,][]{mar71}.
The tables include the values corresponding to the full
set of \citeauthor{nai10} galaxies, as well as those corresponding 
to subsets analyzed below. 
The symbols in Figs.~\ref{nair1}b and \ref{nair1}c
represent $\mu_T(A)$ and $\mu_A(T)$, respectively. 
These figures also include as shaded areas 
the regions around the median containing 68\% (dark gray) 
and 95\% (light gray) of the galaxies.
%
Several results stand out. There is a general
trend for the red galaxies to have early morphological
types, and for the blue galaxies to have late morphological 
types, as quantified in Figs.~\ref{nair1}b and \ref{nair1}c.
However, the relationship has a large scatter with a 
standard deviation between 2 and 3 types, both for the dispersion 
of Hubble types given an ASK class, and for the dispersion of 
spectroscopic classes fixed the Hubble type (see columns labeled 
Stand.~Dev. in Tables~\ref{table_ask} and \ref{table_hubble}).
Figure~\ref{nair4} shows galaxy images that
illustrate the various parts of the scatter plot -- the main 
trend along the diagonal, as well as the outliers represented
by red spirals (upper left) and blue ellipticals (bottom right). 
The distributions of Hubble types given the ASK class are very skewed;
they present extended tails that go to the late morphological 
types for the red galaxies, and to the early morphological
types for the blue spectroscopic classes.
A notable fact is the lack of Sd, Sdm,  Im and Sm  
in ASK classes with red 
spectra. The opposite is not true: ellipticals with emission lines and
very blue spectra are not very common, but they do exist. 
%
%
%
\begin{deluxetable*}{cccc|ccc|ccc|ccc|ccc|ccc}
\tabletypesize{\scriptsize}
\tablewidth{0pt} 
\tablecaption{Statistical parameters characterizing the dispersion 
of Hubble types given an ASK class.\label{table_ask}}
\tablehead{
        \colhead{ASK}                                    &
           \multicolumn{3}{c}{Mean\tablenotemark{d}}                  &
           \multicolumn{3}{c}{Median}& 
           \multicolumn{3}{c}{Mode}                  &
           \multicolumn{3}{c}{Stand. Dev.\tablenotemark{e}}     &             
           \multicolumn{3}{c}{Skewness\tablenotemark{e}}                &  
           \multicolumn{3}{c}{Kurtosis\tablenotemark{e}}\\
\colhead{} &
\colhead{\tablenotemark{a}}&\colhead{\tablenotemark{b}}&\colhead{\tablenotemark{c}}& 
\colhead{\tablenotemark{a}}&\colhead{\tablenotemark{b}}&\colhead{\tablenotemark{c}}& 
\colhead{\tablenotemark{a}}&\colhead{\tablenotemark{b}}&\colhead{\tablenotemark{c}}& 
\colhead{\tablenotemark{a}}&\colhead{\tablenotemark{b}}&\colhead{\tablenotemark{c}}& 
\colhead{\tablenotemark{a}}&\colhead{\tablenotemark{b}}&\colhead{\tablenotemark{c}}&
\colhead{\tablenotemark{a}}&\colhead{\tablenotemark{b}}&\colhead{\tablenotemark{c}}
           }
\startdata
0& S0a$-0.1$&Sa$-0.4$&S0$+ 0.2$& S0a&Sa&ES0&E&E&E&2.6&2.7&2.4&0.3&0.1&0.8&
-1.2&-1.2&-0.8\\
1&Sa$+ 0.5$& Sab$+0.4$&  S0$+ 0.3$& Sab&Sb&   ES0& Sb&    Sb&       E& 2.5&   
2.3&   2.4& -0.4&  -0.6&   1.0& -1.0&  -0.2&  -0.1\\ 
2&  S0$+0.2$& S0a$-0.5$&  S0$-0.1$& S0&    S0&   ES0& E&     E&       E&
2.4&   2.2&   2.3& 0.8&   0.7&   1.0& -0.6&  -0.2&  -0.2\\
3&  Sa$-0.4$&Sa$-0.4$& S0a$+ 0.4$& S0a&   S0a&   S0a& E&   ES0&       E&
 2.7&   2.8&   2.7& 0.3&   0.5&   0.4& -1.2&  -1.1&  -1.2\\
4&  Sb$-0.0$&  Sb$+0.2$& Sab$+ 0.3$& Sb&   Sbc&   Sb& Sc&    Sc&      Sb&
 2.0&   2.1&   2.1& -0.7&  -0.6&  -0.4& 0.1&  -0.5&   0.0\\
5&  Sb$-0.1$&  Sb$+ 0.1$&  Sb$-0.4$& Sb&   Sbc&   Sb& Sb&    Sc&      Sb&
2.0&   2.2&   2.0& -0.8&  -0.8&  -0.8& 0.0&  -0.2&  -0.1\\
6& Sab$+0.1$&Sab$-0.2$& Sab$-0.1$& Sb&   Sab&   Sb& Sb&    Sa&     Sb&
 2.0&   2.0&   2.0& -0.6&  -0.2&  -0.6& -0.5&  -0.9&  -0.5\\
7& Sab$-0.4$&Sa$-0.4$& Sab$-0.4$& Sa&   S0a&   Sa& S0a&   S0a&     S0a&
 2.0&   1.6&   2.0& 0.2&   1.2&   0.1& -1.1&   1.4&  -1.1\\
8&Sab$-0.3$&--- & Sab$+ 0.3$& Sa&---&   
   Sa& Sa&---&      Sa& 1.9&---&   1.9& 
   1.0&---&   0.7& -0.8&---&  -1.5\\
9& Sbc$-0.3$& Sbc$+ 0.4$&  Sb$+0.5$& Sbc&    Sc&   Sbc& Sc&    Sc&
      Sc& 2.0&   2.0&   2.0& -1.1&  -1.4&  -1.1& 1.2&   2.3&   1.0\\ 
10&  Sb$+0.3$&  Sb$+0.5$&  Sb$-0.1$& Sbc&   Sbc&   Sb& Sc&    Sc&
      Sb& 2.1&   2.4&   2.0& -0.5&  -0.6&  -0.5& 0.3&   0.1&  0.3 \\
11&  Sb$-0.1$&  Sb$-0.3$&  Sb$-0.4$& Sb&    Sb&   Sb& Sc&    Sc&
      Sb& 2.1&   2.7&   2.0& -0.2&   0.0&  -0.3& 0.5&   0.1&0.3\\
12& Sbc$+0.2$&  Sc$+ 0.1$& Sbc$-0.1$& Sc&   Scd&   Sbc& Sc&   Scd&
      Sc& 2.1&   1.8&   2.1& -0.9&  -1.5&  -0.7& 0.8&   2.5&   0.5\\
13&  Sc$+0.1$& Scd$-0.2$&  Sc$-0.2$& Scd&   Scd&   Scd& Scd&   Scd&
     Scd& 2.5&   1.9&   2.7& -1.4&  -1.9&  -1.2& 1.9&   6.4&   1.1\\
14& Sbc$+ 0.1$& Sbc$+ 0.1$& Sbc$-0.2$& Sc&    Sc&   Sbc& Sc&    
Sc&      Sc& 2.5&   2.9&   2.6& -0.6&  -0.6&  -0.5& 0.6&   0.1&   0.4\\
16& Scd$-0.0$& Scd$+0.4$& Scd$-0.3$& Scd&    Sd&   Scd& Scd&    Sd&
     Scd& 2.4&   2.5&   2.6& -1.1&  -1.2&  -1.0& 1.9&   1.7&   1.1\\
18& Sbc$+ 0.0$&Sbc$-0.2$&  Sb$+ 0.5$& Sc&   Sbc&   Sbc& Sc&    Sd&
      Sc& 3.1&   3.1&   3.3& -0.4&  -0.3&  -0.2& -0.7&  -1.0&  -0.9\\
19& Scd$+ 0.1$&  Sd$-0.4$& Scd$-0.4$& Sd&    Sd&   Sd& Scd&   Scd&
     Sm& 2.9&   2.5&   3.5& -1.0&  -1.2&  -0.8& 0.4&   1.5&  -0.6\\ 
22& Scd$-0.1$& Scd$-0.4$& Sbc$+ 0.3$& Sd&   Scd&   Scd& Sd&   Scd&
       E& 3.9&   3.6&   4.6& -1.0&  -0.7&  -0.3& -0.0&  -0.8&  -1.4\\ %
23&  Sd$+0.1$&  Sd$-0.2$&  Sd$-0.2$& Sm&   Sdm&   Sm& Sm&    Sm&
      Sm& 3.5&   3.2&   4.1& -1.6&  -1.4&  -1.3& 1.2&   0.9&   0.1\\ 
26&  Sc$+ 0.0$&  Sc$+0.1$& --- &
  Scd&   Scd&---&
  Sa&   Scd&---&
  2.9&   2.8&--- &
  -0.5&  -0.6&---&
   -1.5&  -1.3&---
\enddata
\tablecomments{Morphological types from \citet{nai10}.
         ASK classes not included imply having no enough galaxies in 
        the bin to compute statistical properties. 
        Hubble~types are coded in Table~\ref{equiv_morpho}}.
\tablenotetext{a}{Full set of galaxies}
\tablenotetext{b}{Full set, V$_{\rm max}$ corrected}
\tablenotetext{c}{$b/a > 0.5$, to avoid edge-on systems}
\tablenotetext{d}{Hubble~type $\pm$ fraction of Hubble~type}
\tablenotetext{e}{In units of Hubble~type}
\tablenotetext{------}{{\bf To be printed in portrait mode}}
%
\end{deluxetable*}
%
%
\begin{deluxetable*}{cccc|ccc|ccc|ccc|ccc|ccc}
\tabletypesize{\scriptsize}
\tablecolumns{19}
\tablewidth{0pt} 
\tablecaption{Statistical parameters characterizing the dispersion 
of ASK classes once the Hubble type is fixed.\label{table_hubble}}
\tablehead{
        \colhead{Hubble Type\tablenotemark{d}}                                  
 &
           \multicolumn{3}{c}{Mean}                  &
           \multicolumn{3}{c}{Median}& 
           \multicolumn{3}{c}{Mode}                  &
           \multicolumn{3}{c}{Stand. Dev.}     &             
           \multicolumn{3}{c}{Skewness}                &  
           \multicolumn{3}{c}{Kurtosis}\\
           \colhead{ } &
\colhead{\tablenotemark{a}}&\colhead{\tablenotemark{b}}&\colhead{\tablenotemark{
c}}&
\colhead{\tablenotemark{a}}&\colhead{\tablenotemark{b}}&\colhead{\tablenotemark{
c}}& 
\colhead{\tablenotemark{a}}&\colhead{\tablenotemark{b}}&\colhead{\tablenotemark{
c}}& 
\colhead{\tablenotemark{a}}&\colhead{\tablenotemark{b}}&\colhead{\tablenotemark{
c}}& 
\colhead{\tablenotemark{a}}&\colhead{\tablenotemark{b}}&\colhead{\tablenotemark{
c}}&
\colhead{\tablenotemark{a}}&\colhead{\tablenotemark{b}}&\colhead{\tablenotemark{
c}}
           }
\startdata
E&   1.5&   2.3&1.5& 2&   2&   2& 2&      2&      2& 1.4&   2.7&   1.4& 2.7&
   2.7&   2.7& 21.0&   8.1&21.2\\ 
ES0&   2.1&   3.0&   2.1& 2&   2&   2& 2&      2&      2& 2.2&   2.8&   
2.2& 2.4&   2.4&   2.4& 8.1&   5.6&   8.2\\
S0&   2.4&   3.2&   2.5& 2&   2&   2& 2&      2&      2& 2.5&   3.2&
   2.6& 2.2&   1.9&   2.1& 6.1&3.4&   5.5\\
S0a& 2.6&   4.2&   2.9& 2&   3&   2& 2&      2&      2& 2.9&   4.2&
   3.1& 1.7&   1.31&   1.6& 3.1&   0.7&   2.3\\ 
Sa&   2.9&   5.0&   3.6& 2&   3&   2& 0&      2&      2& 3.1&   4.6&
   3.3& 1.5&   0.9&   1.1& 2.0&  -0.4&   0.8\\ 
Sab&   3.0&   4.3&   3.6& 2&   3&   2& 0&      0&      2& 3.1&
   3.8&   3.2& 1.1&   0.8&   0.9& 0.6&  -0.1&  -0.0\\ 
Sb&   3.8&   5.2& 4.6& 3&   5&   5& 0&      0&      2& 3.4&
   4.0&   3.5& 0.8&   0.5&   0.5& -0.2&  -0.9&  -0.5\\ 
Sbc&   4.8&   6.3&   5.4& 5&   9&   5& 5&      9&      5& 3.4&   
4.0&   3.3& 0.4&   0.2&   0.2& -0.5&  -0.9&  -0.6\\ 
Sc&   6.1&   9.1&  8.8& 9&   9&   9& 9&      9&      9&
    3.5&   3.8&   3.3& 0.1&  -0.1&  -0.1& -0.7&  -0.8&  -0.6\\ 
Scd&  11.9&  12.4&  12.2& 12&   13&   13& 13&     13&     13& 2.8&
   2.6&   2.5& -0.6&  -0.6&  -0.8& 0.1&   0.9&   0.8\\ 
Sd& 13.4&  13.8&  13.3& 13&   14&   13& 13&     13&     13& 2.5&
   2.4&   2.5& -0.9&  -1.2&  -1.3&   1.8&   3.0&   3.2\\ 
Sdm&  14.2&  15.7& 14.0& 14&   14&   14& 13&     13&     13&
    2.0&   2.0&   1.9& -0.0&   0.2&   0.0& -0.7&  -1.1&  -0.4\\ 
Sm&  15.6&  16.1&  15.5& 16&   16&   16& 16&     16&     16&
 2.0&   1.7&   2.0& -1.3&  -0.5&  -1.3& 4.3&   0.2&   4.3\\ 
Im&  15.7&  16.2&  14.2& 16&   16&   16& 16&     16&     
16& 2.5&   2.1&   2.6& -0.7&  -0.8&  -0.6& -0.4&   0.9&  -0.5 
\enddata

\tablecomments{Morphological types from \citet{nai10}. Only major ASK classes are
considered to compute the statistical parameters.}
\tablenotetext{a}{Full set of galaxies}
\tablenotetext{b}{Full set, V$_{\rm max}$ corrected}
\tablenotetext{c}{$b/a > 0.5$, to avoid edge-on systems}
\tablenotetext{d}{As coded in Table~\ref{equiv_morpho}}
\tablenotetext{------}{{\bf To be printed in portrait mode}}
%
%
%
\end{deluxetable*}

%
%
\begin{figure*}
\centering
\includegraphics[width=0.6\textwidth,angle=90]{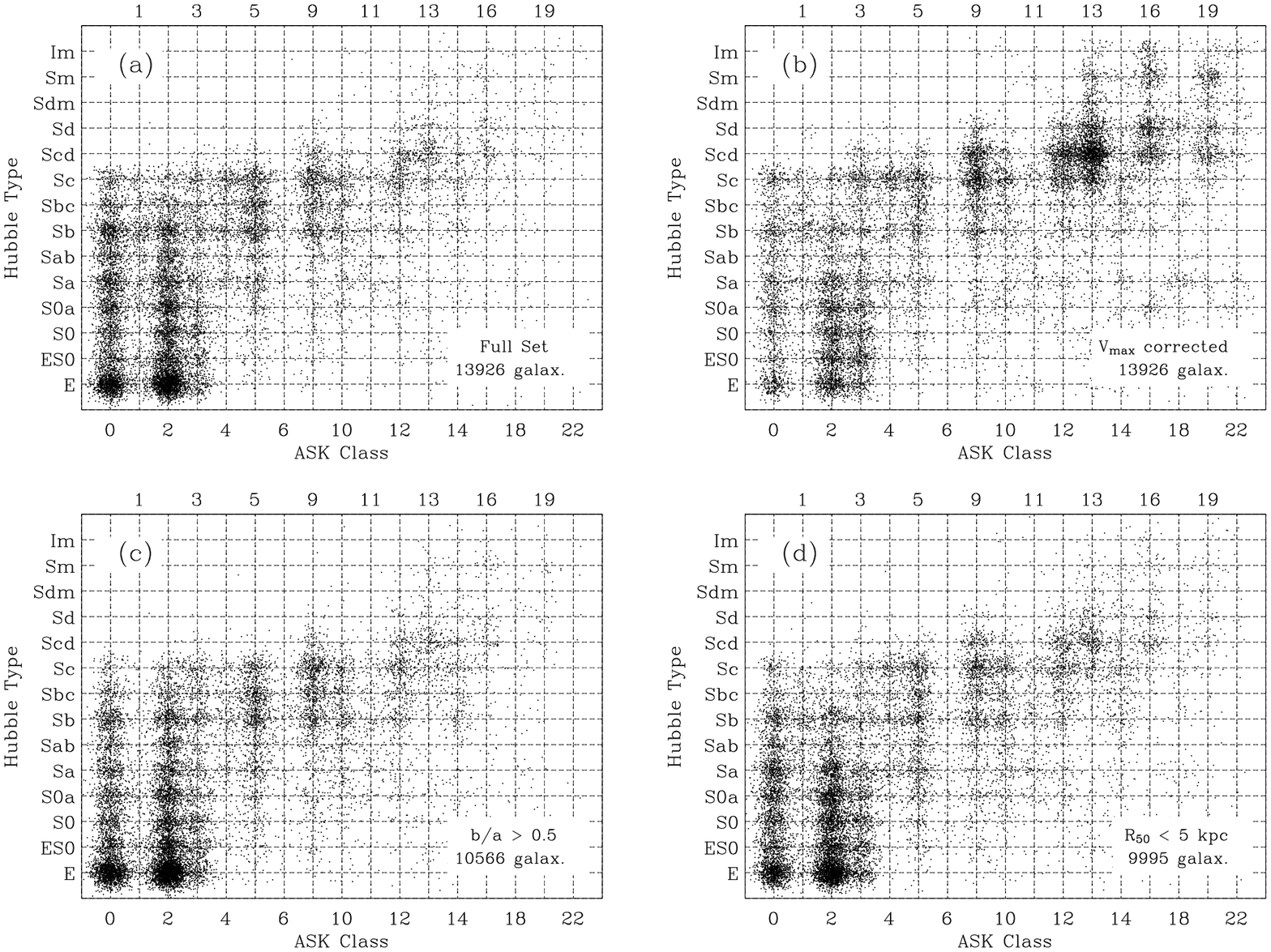}\\
\includegraphics[width=0.6\textwidth,angle=90]{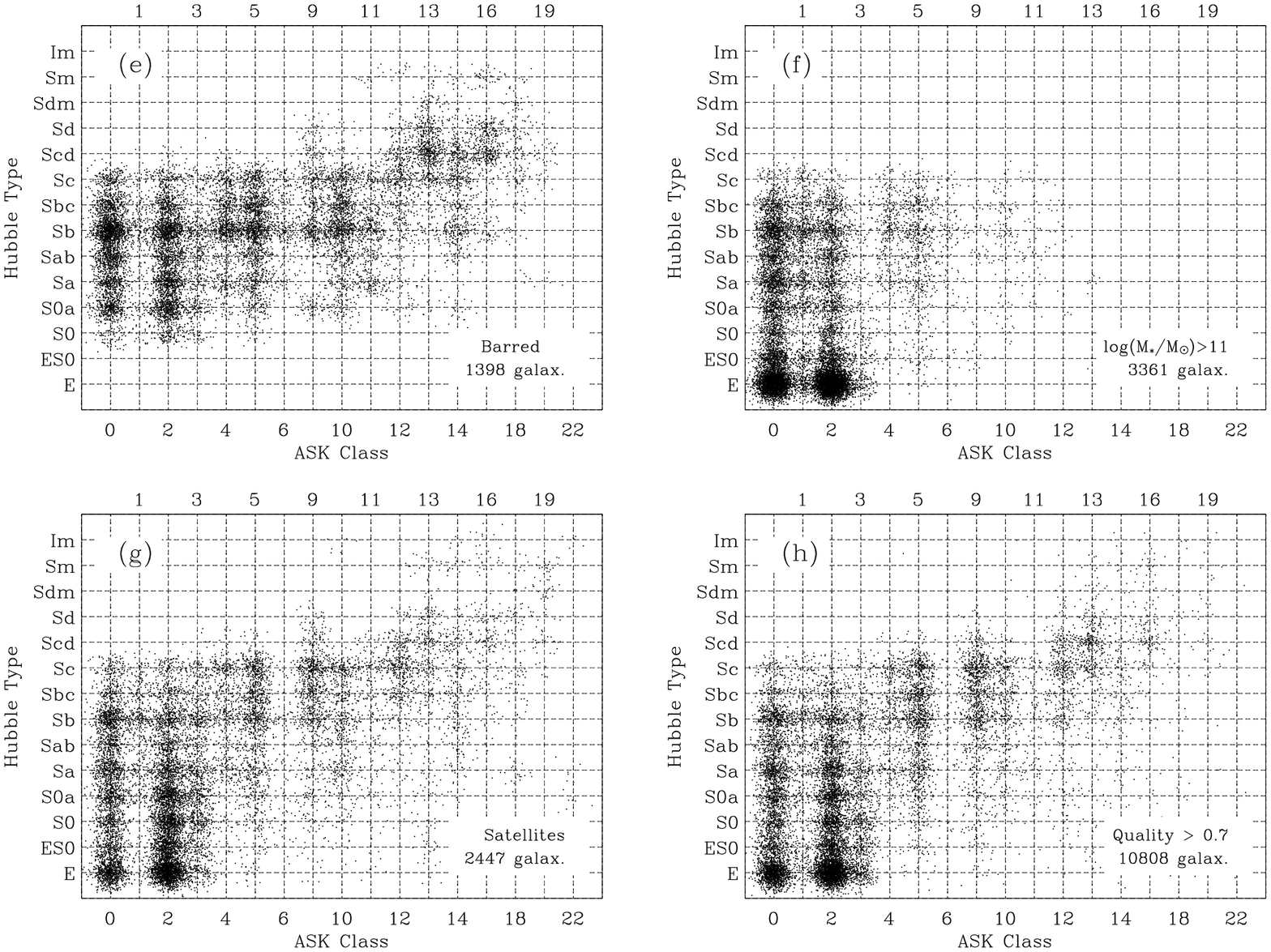}
\caption{
\scriptsize
Scatter plots of Hubble type versus ASK class for galaxies in \citet{nai10}.
(a) Same as Fig.~\ref{nair1}a.
All the other panels represent trims on this galaxy set.
(b) Full set corrected so that, rather than a magnitude limited sample,
it represents a volume limited sample.
(c) Only rounded galaxies are included, which removes edge-on
disk galaxies. See how ASK~1 and 4 have been greatly 
reduced as compared to (a). 
(d) Small galaxies, with half-light radius $R_{50} < 5~$kpc.
(e) Only galaxies labeled as barred. 
(f) High mass galaxies (stellar mass $M_\star > 10^{11}\,M_\odot$). 
Blue types are gone, ellipticals are greatly reinforced,
and the relative number of ASK~0 galaxies has also increased.
(g) Satellite galaxies. They are tracing high density environments. 
(h) Only galaxies with high confidence ASK classification.
All plots contain the same number of
points, but the insets at the bottom right corners give the number of 
galaxies left by the particular selection
(see the main text for details).
Only major ASK classes are included.
}
\label{nair2}
\end{figure*}
\begin{figure}
\includegraphics[width=0.5\textwidth]{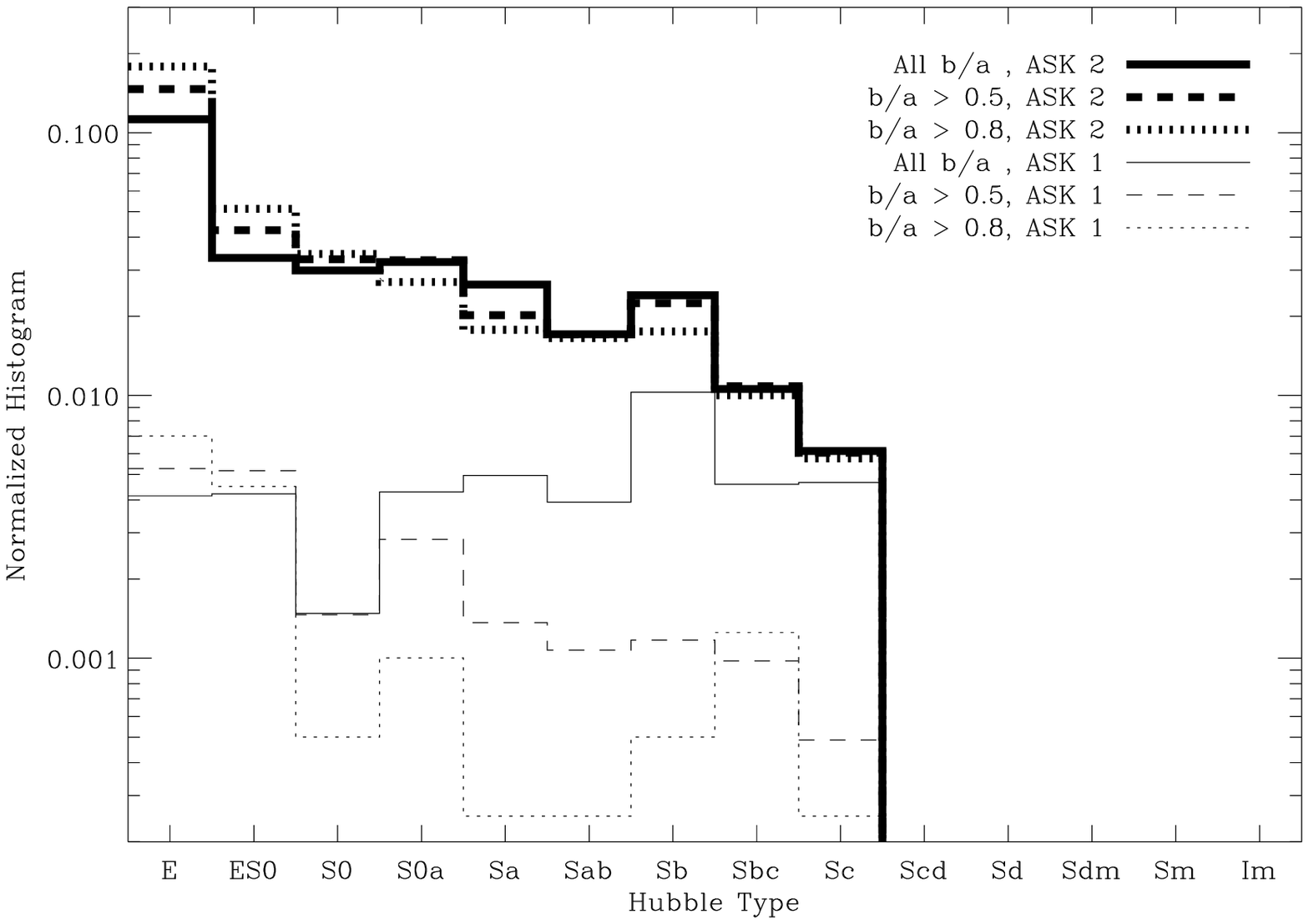}
\caption{
Histograms of Hubble types for galaxies of different ellipticities $b/a$; 
from the full range of possible values ($b/a > 0,$ the solid lines)
to more rounded galaxies ($b/a > 0.5$, the dashed lines, and $b/a > 0.8$, 
the dotted lines). The thick and thin lines represent 
ASK~2 and ASK~1 galaxies, respectively. Note how ASK~1 galaxies 
tend to disappear with increasing
$b/a$, meaning that they mostly are edge-on (disk) systems. 
The axis-ratio threshold has almost no effect on ASK~2, 
except for a slight increase of early types
when considering rounded galaxies. 
The histograms are normalized to the total number of galaxies,
i.e., galaxies of all ASK classes and Hubble types.}
\label{new_fig1}
\end{figure}
\begin{figure}
        \centering
        \includegraphics[width=.5\textwidth,angle=0]{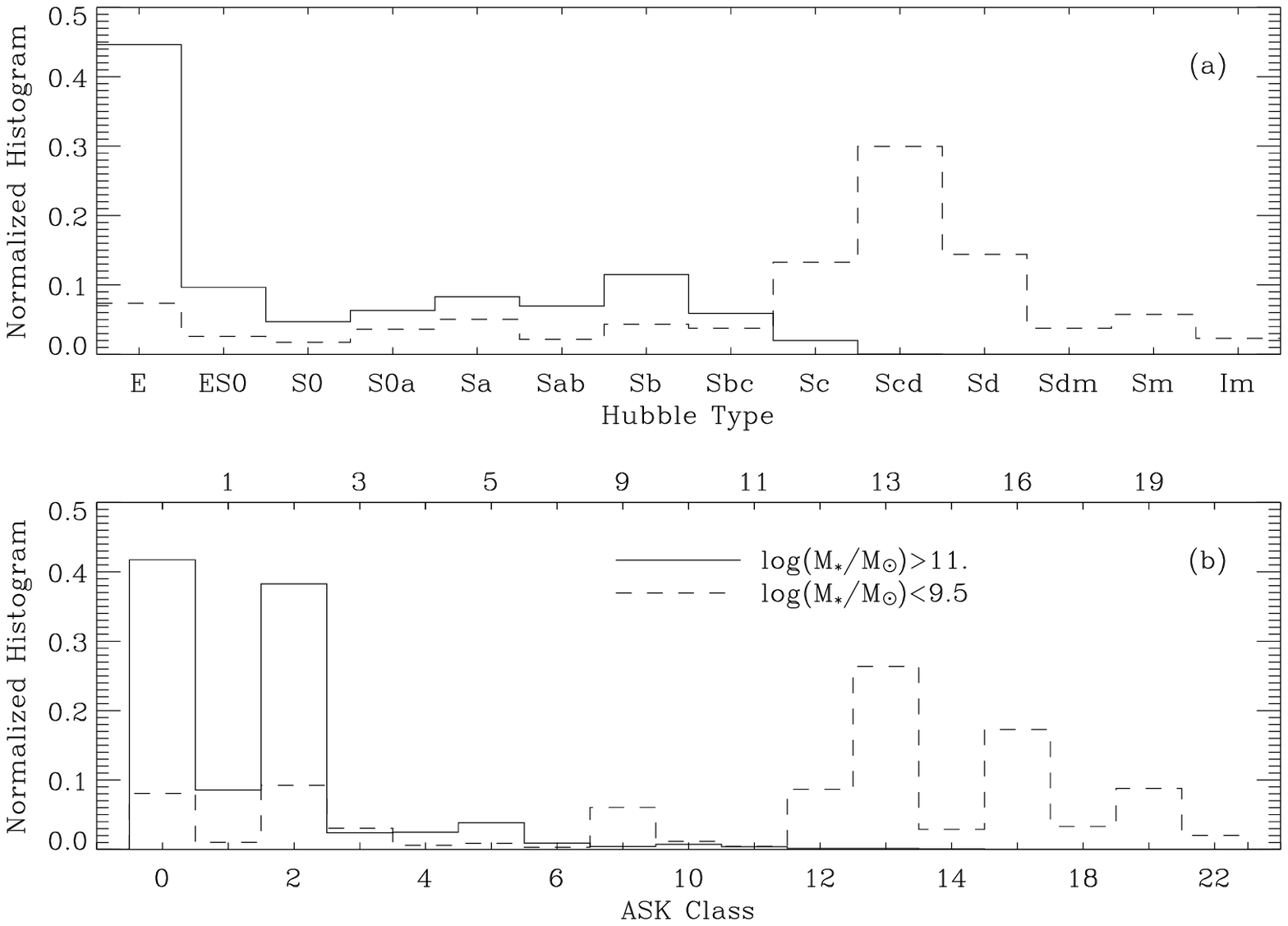}
        \caption{
        (a) Distribution of Hubble types among the massive galaxies 
        (stellar mass  $M_\star > 10^{11}\,M_\odot$, the solid line) and 
        the low mass galaxies ($M_\star < 10^{9.5}\,M_\odot$, the dashed
        line) in the catalog by \citet{nai10}.
        (b) Distribution of ASK classes for the same two sets of galaxies.
        All histograms are normalized to one.
        }
        \label{referee2}
        \end{figure}

The scatter shown in Figs.~\ref{nair1}, and quantified
in Tables~\ref{table_ask} and \ref{table_hubble}, 
has been inferred from the particular sample of galaxies selected by \citet{nai10}.
However, with small modifications, it is representative of
the relationship and scatter existing among local galaxies. 
One can support this claim by comparing the results
with those obtained using other morphological catalogs,
an approach to be pursued in the forthcoming subsections.
In addition, the catalog by \citet{nai10} is large enough
to be partaken, thus allowing the analysis of whether
the relationship depends on particular properties of the  galaxy sample. 
The rest of the section is devoted to such exercise. We have constructed scatter plots 
Hubble-type-vs-ASK-class using subsets of \citeauthor{nai10} galaxies.
Some of them are shown in Fig.~\ref{nair2}.
Obviously, re-sampling the original set reduces the
number of galaxies, which makes it difficult to visually
compare different scatter plots\footnote{The 
number density of points on the plot is integrated by 
the  brain when visualizing the histogram. However, the
local number density scales with the total
number of represented galaxies, a spurious dependence 
that should be removed to allow the direct comparison 
between subsets having different number of galaxies.}.
In order avoid this unwanted dependence, we show the same
number of galaxies in all plots. They are
not real galaxies but points randomly 
drawn from the histograms inferred from the real galaxies.
The number of real galaxies left by the selection is
given in the inset of the corresponding panel. 
Figure~\ref{nair2}a shows the scatter plot of the original 
distribution and it is included for reference. 
(It is equivalent to Fig.~\ref{nair1}a.)  
Figure~\ref{nair2}b shows the scatter plot in the 
case that the sample were volume limited,
i.e., if it would include all galaxies within
a fixed volume. We have used the $V_{\rm max}$ approach
to construct the sample, as explained in appendix~\ref{appa}. 
The importance of the blue galaxies has increased 
with respect to Fig.~\ref{nair2}a because intrinsically faint
galaxies are underrepresented in magnitude limited
samples (Malmquist bias), and 
many of them tend to be blue \citep[e.g.,][]{bal04,bla09}.
However, the original dispersion remains. This fact is 
quantitatively corroborated in Tables~\ref{table_ask} and
\ref{table_hubble}; compare the standard deviations
in columns $(a)$ and $(b)$. 
Figure~\ref{nair2}c shows the scatter plot for
rounded targets, which excludes edge-on disk galaxies.
Explicitly, we show galaxies where the ratio between the minor 
and major axes $b/a > 0.5$. 
The scatter remains, although the density of red late type
galaxies is somehow reduced. 
This can be better appreciated in Fig.~\ref{new_fig1}, the thick lines,
which portrays histograms of ASK~2 galaxies for various 
aspect ratios -- the shape of the distribution remains, but with
a slight relative increase of E-S0 galaxies with increasing $b/a$ threshold.
Part of the red spirals are edge-on 
galaxies, which are known to be reddened with
respect to their face-on counterparts
\citep[e.g., ][and \S~\ref{discussion}]{gio94,mas10b}.
However, not all the
red spirals are edge-on systems (see Figs.~\ref{nair4},
\ref{nair2}c and \ref{new_fig1},
as well as \citeauthor{mas10}~\citeyear{mas10}). Note also
that most galaxies corresponding to ASK~1 and 4 have
disappeared from  Fig.~\ref{nair2}c, indicating that
these classes seem to be preferentially formed by edge-on spirals.
This drastic decrease is also illustrated in Fig.~\ref{new_fig1},
where the thin lines show how the relative number 
of ASK~1 galaxies drops down when selecting 
only rounded galaxies. (ASK~4 galaxies behave similarly.)
The columns $(c)$ in Tables~\ref{table_ask} and \ref{table_hubble}
show the statistical parameters for these face-on systems,
which are not very different from those of the original sample,
listed in columns $(a)$.
Figure~\ref{nair2}d shows the scatter plot considering only 
small galaxies, namely galaxies with half-light
radius $R_{50} < 5\,$kpc. SDSS spectra are obtained
with a 3\,arcsec fiber,  which covers only the central
parts of the large galaxies.
Statistically, small galaxies look small on the sky, therefore, the
fact that small galaxies show a scatter similar to the full sample
indicates that red spirals are 
not artificially produced by the fiber covering only the central (red) 
bulge, since this effect would preferentially affect
large galaxies.
Obviously, the above argument would be stronger using angular sizes rather 
than physical sizes, but we consider it sufficient because comparisons 
made with the other morphological catalogs using the proper angular 
sizes also discard the bias (see \S~\ref{morph_fuk} and \ref{quantify}).
%
%
Figure~\ref{nair2}e shows the scatter plot for
barred galaxies \citep[i.e., flagged as strong or intermediate barred
in][]{nai10}. 
Ellipticals have disappeared since bars are associated with disks in 
galaxies \citep[e.g.,][]{men10}, and this lack of ellipticals
reduces the scatter of the plot in the region
ASK 0--3. Barred S0 galaxies are also rare in Fig.~\ref{nair2}e,
which was expected in view of the known shortage 
of bars in S0s  \citep[see, e.g.,][]{bar08,agu09,but10}.  
%
Figure~\ref{nair2}f shows the scatter plot for massive galaxies 
with stellar masses $M_\star > 10^{11}\,M_\odot$. Only red classes 
remain. (Massive galaxies tend to be red, even if 
they are spirals; see, \citeauthor{bla09}~\citeyear{bla09} and 
references therein.)
ASK~0 results particularly enhanced independently of whether
they are elliptical or spirals. Note that ASK~12 and bluer classes 
do not seem to be associated with massive galaxies.
All these properties are more clearly illustrated by the solid lines 
in Fig.~\ref{referee2}, which are just projections
of Fig.~\ref{nair2}f in ordinates (Fig.~\ref{referee2}a)
and abscissae (Fig.~\ref{referee2}b). For comparison,  
Fig.~\ref{referee2} also shows the histograms for 
low mass galaxies ($M_\star < 10^{9.5}\,M_\odot$; the dashed lines). 
In contrast with high mass galaxies, low mass galaxies 
tend to be blue  (ASK~12 and larger), and have late Hubble types 
(Sc and later), although, low mass galaxies of all colors and 
morphologies exist.
The catalog by \citet{nai10} contains a flag 
indicating if the galaxy is or not the most luminous of its group.
Figure~\ref{nair2}g represents those that are not, i.e.,
those marked as satellite galaxies. The scatter remains indicating
that the presence of companion galaxies (and thus of interaction)
does not significantly modifies the diagram. However, there
is a subtle difference with respect to Fig.~\ref{nair2}a, 
namely, an excess of S0 and latter types with respect
to the ellipticals. This preference is also known to be 
attributable to high density environments.   
Finally, Fig.~\ref{nair2}h shows that the scatter is not produced 
by spectral misclassifications. It shows galaxies with 
high ASK {\em quality}, specifically, with a probability of belonging 
to its class larger than 70\% -- see \citet{san10} for
details on the definition of quality.
This high quality subset represents 78\% of the 
\citeauthor{nai10} sample.

\begin{figure}
\centering
\includegraphics[width=.5\textwidth,angle=0]{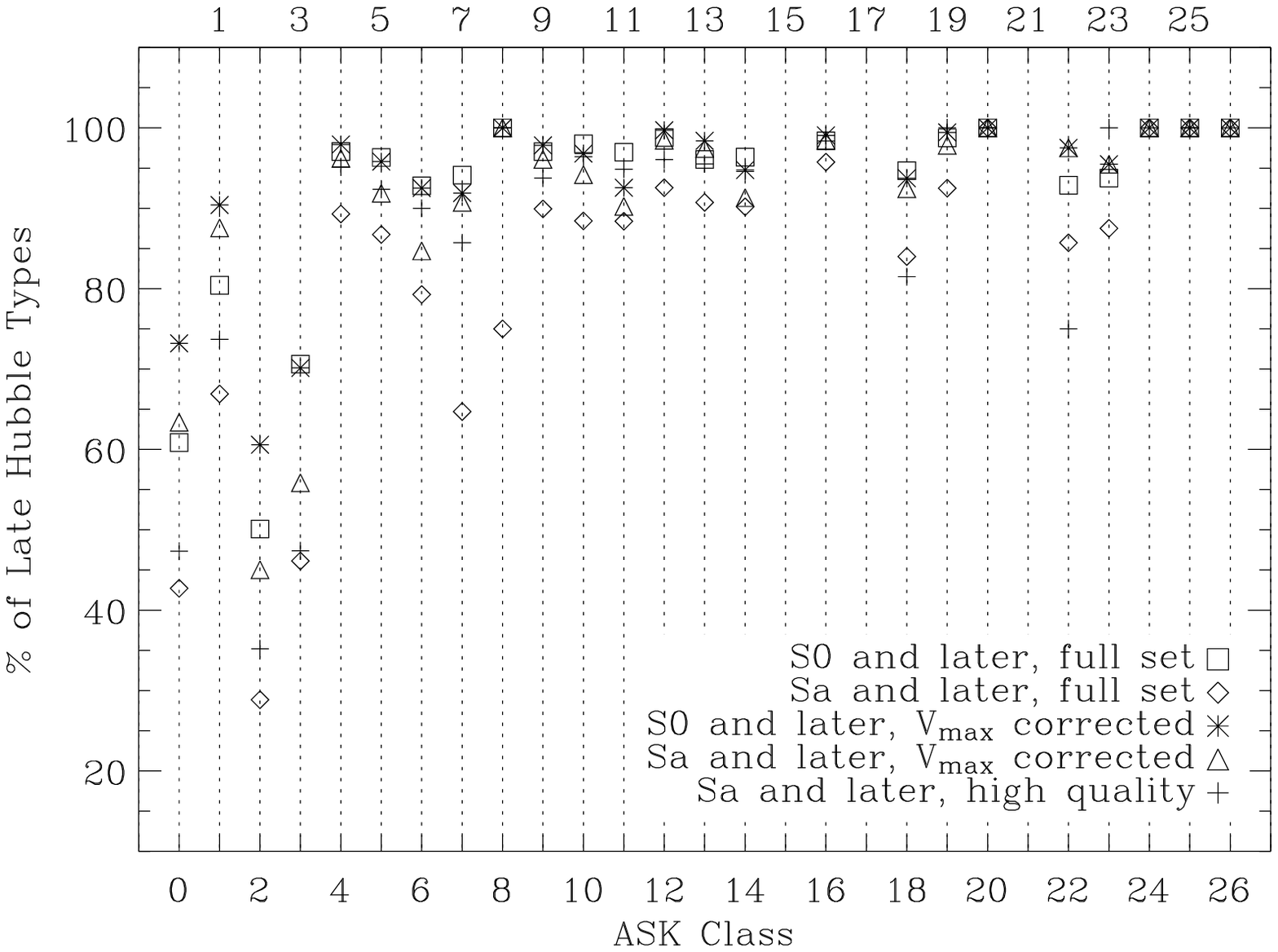}
\caption{
Percentage of late Hubble types corresponding to each ASK class.
Late types dominate the blue classes, but they are also
very common among red spectra.
The different symbols represent various galaxy
samples, as indicated in the inset. The squares
correspond to the full \citet{nai10} data set
when late type means S0 and later. The rhombi also 
correspond to the same original data for Sa  galaxies and later.
The asterisks and triangles consider a volume limited sample. 
Finally, the plus signs correspond to the original  data set
removing galaxies with uncertain ASK classes.  
}
\label{sp_vs_ask}
\end{figure}

Figure~\ref{sp_vs_ask} contains 
the percentage of late type galaxies
in each ASK class. It shows how spirals are present in
all spectral classes. Obviously they dominate the blue
ASK classes (4 and bluer), but their contribution 
is not negligible even in the red ASK~0, 2 and 3 that
are characteristic of ellipticals. For example, in the
original sample of \citet{nai10}, 
57\% of the red ASK~0, 2 and 3 are formed by spirals (S0 and later).
The actual fraction depends on the used sample
(see Fig.~\ref{sp_vs_ask}),  but red spirals turn out to be common. 

Based on Figs.~\ref{nair1}, \ref{nair4},
\ref{nair2}, \ref{new_fig1}, \ref{referee2}, 
and \ref{sp_vs_ask}, Tables~\ref{table_ask} and
\ref{table_hubble}, as well as on additional explorations, we distill the 
following results (sorted not necessarily according to importance):
\begin{itemize}
\item[-] We recover the well known global trend indicating that
 early Hubble types tend to be red and late Hubble types
 tend to be blue. This trend presents a large scatter
 which has been quantified in Tables~\ref{table_ask} and
\ref{table_hubble}.  Given a Hubble type, the ASK class 
varies with a standard deviation between 2 and 3 classes. 
The same happens with the Hubble type once the ASK class is fixed. 
 \item[-]
 The scatter is real (see Fig.~\ref{nair4}). It is not produced
 by problems in the classification, and it is not reduced
 when particular subsets are considered -- low and high galaxy masses, 
 low and high density environments, barred and non-barred galaxies,
 face-on galaxies, small and large galaxies, or even when a volume limited
 sample is considered. 
\item[-] The upper left corner of the scatter plots is empty. 
There are no Sd, Sdm, Sm or Im galaxies with red spectra. 
This seems to indicate that 
morphological evolution is somewhat
faster than color evolution
for these morphological types
(commonly associated with dwarf galaxies).
\item[-] On the contrary, there are lots of
spiral galaxies in red ASK classes,
suggesting that the spectrum can
redden without morphological transformation,
provided that the galaxy is massive enough.
We find that 57\% of the red ASK~0, 2 and 3 galaxies are not 
ellipticals (i.e., are neither E nor ES0).
\item[-] Even though red spectra are not associated
with ellipticals, most ellipticals do have red
spectra: 93\% of the ellipticals in 
\citet{nai10} belong to ASK~0, 2 or 3.   
\item [-] There is also a 3\% of 
blue ellipticals (see Fig.~\ref{nair4}), like those studied by 
\citet[][see also the references in \S~\ref{introduction}]{hue10a}.
\item[-]
Low mass dwarf galaxies tend to be  late type (Sc
 and later) and blue (ASK~12 and  bluer). 
\item[-]ASK  1 and 4 seem to be made of edge-on (reddened) 
        spirals.
\item[-] All red classes contain edge-on spirals, however, not all
red spirals are edge-on. The reddening associated with being
edge-on may account for some of them, but not for all. 
\item [-] S0s tend to be satellites, 
        i.e., they are not the most luminous of their groups.
\item [-] Bars are not present in Es, and
        are almost absent in S0s.
\item [-]
        ASK~6 is made out of active galactic nuclei \citep[AGN; ][]{san10},
        and we find it to have a particularly 
        flat distribution of Hubble types, going all the way from 
        E to Sc, but not later 
        (see Figs.~\ref{nair2}a and \ref{nair2}b).
%
\item[-]The galaxies with transition colors 
        are preferentially early spirals. As it is shown in 
	\citet{san10}, ASK~5 is formed by galaxies
        right in the green valley and, according 
        to Fig.~\ref{nair2}a,  87\% of them are Sa or later 
        types. (96\% are S0 or later.) 
\end{itemize} 

%
\subsection{Hubble types derived by \citet{fuk07}}
\label{morph_fuk}

\begin{figure}
\centering
\includegraphics[width=0.5\textwidth]{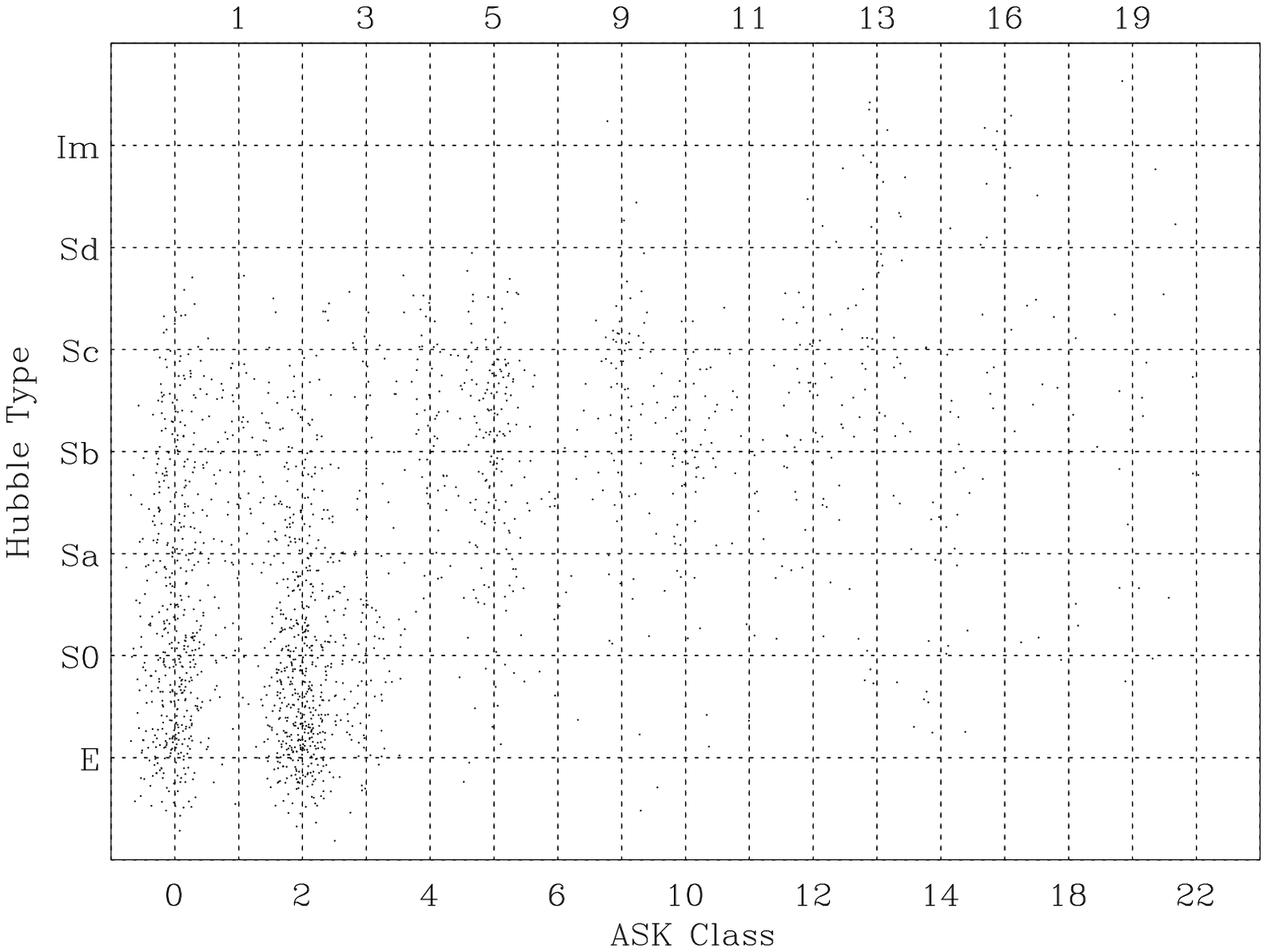}
\caption{
Scatter plot of Hubble type vs ASK class for the
SDSS/DR3 galaxies classified by \citet{fuk07}.
In order to avoid the overlapping of 
galaxies with the same Hubble type and ASK class, we
have added small artificial random shifts to all points.
It is very similar to Fig.~\ref{nair2}a, except that
this sample contains only one tenth of the galaxies
represented therein.
Only major ASK classes are included.
}
\label{ask_vs_morph}
\end{figure}

\citet{fuk07} carried out a morphological classification of bright galaxies 
in the north equatorial stripe of SDSS/DR3. The classification has been 
performed by visual inspection of SDSS~$g$ images by three different observers. 
The catalog contains 2253 galaxies, but only 1866 galaxies have 
spectra, and so, ASK class. The classification comprises 
only six Hubble types (see Table~\ref{equiv_morpho}), but
the fact that it 
has been agreed by three qualified observed makes it very reliable
-- the mean standard deviation among the three classifications is 
only 0.4 types.
The resulting scatter plot morpho-type vs spectro-class 
is presented in Fig.~\ref{ask_vs_morph}. 
There is no significant difference when compared to
Fig.~\ref{nair2}a. The same global trends, including
the presence of red spirals and blue ellipticals, and
the absence of red Sd---Im. This sample also allowed us 
to check that the spirals in red ASK classes
remain in place even when considering
rounded targets and small ($R_{50} < 3$~arcsec) galaxies.
Moreover, the plot is insensitive to considering or not galaxies
flagged in the catalog as peculiar.

%
%

\subsection{Morphology from the 3rd reference catalog, RC3, by \citet{dev91}}
\label{sec_rc3}

The 3rd reference catalog of bright galaxies (RC3) by \citet[][]{dev91} has 
been the standard for morphological classifications during the
last two decades, therefore we also wanted to check the above
results against this reference. For this reason, and also
because their classification is finner and more precise
than the others, we use it as a guide for the equivalence between 
different schemes (Table~\ref{equiv_morpho}). 
The morphological types of RC3 are coded with $T$~types spanning 
from $-6$ to $+ 11$. 
Paraphrasing \citet{dev94}, the catalog includes $T$~types along 
the extended Hubble sequence  through the four main classes 
E (ellipticals or spheroidals,  encompassing stages T~$=-6$ to T~$= -4$, 
from compact to late), 
L (lenticulars or S0s, i.e. arm-less disks, T~$= -3$ to T~$= -1$, from early to
late), 
Sp (spirals from S0a at T~$= 0$ to Sm at T~$= 9$ through the familiar stages 
    a, b, c, d at T~$= 1, 3, 5, 7$ 
    with the transition types ab, bc, cd, dm at T~$= 2, 4, 6, 8$), and
Im (Magellanic irregulars, T~$= 10$, with 11 for compact).

\begin{figure}
\centering
\includegraphics[width=0.5\textwidth]{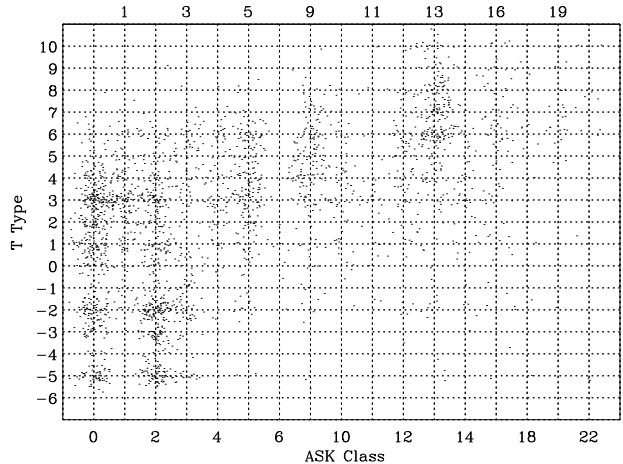}
\caption{
Scatter plot of $T$~type, as given in the RC3 catalog, vs ASK class.
Compare it with the reference Fig.~\ref{nair2}a.
This plot contains all the 3026 galaxies in the match we carry out.
Only major ASK classes are included.
}
\label{ask_vs_rc3a}
\end{figure}

The catalogs discussed so far are based on SDSS images,
therefore, matching them with the ASK classification was trivial
since SDSS targets have unique identifiers. However, this is
not the case for RC3. The version of RC3 included in SDSS/DR7
just contains the right ascension  and declination of the targets. 
In order to carry out the match, we search SDSS/DR7 for galaxies with 
spectra within 30~arcsec of the RC3 coordinates.
We handled multiple matches by choosing the closest object.
The RC3 catalog in SDSS/DR7 contains 23011 entries, with 
only 17801 having valid $T$~type. 
From them we match 3026 galaxies, which is roughly consistent 
with the  solid angled covered by SDSS/DR7 assuming the RC3 
galaxies to be evenly spread throughout the sky.
The scatter plot $T$~type vs ASK~class corresponding to these 
targets is shown in Fig.~\ref{ask_vs_rc3a}. It displays
all the features already described in \S~\ref{sect_nair_abra}, and 
we will not repeat them here. We note, however,
how the presence of red spirals is even more evident 
than in Fig.~\ref{nair2}a.
This increase may be a bias specific to the RC3 catalog, which
contains nearby galaxies, and so, galaxies of large angular size 
(mean half-light radius of 10\arcsec$\pm 5$\arcsec). Given the
finite size of the fiber feeding the SDSS spectrograph, the bulge of the 
spirals contribute more with increasing apparent size, which reddens 
the SDSS spectra.

\subsection{Morphology using galSVM by \citet{hue11}}
        \label{svm_sect}
        
The support vector machine procedure galSVM
by \citet{hue08} is a machine learning algorithm
which tries to find the optimal boundary between
clouds of points in a high-dimensional
space, even when the borders are not linear.
It does not deliver a binary classification but the probability 
of belonging to a given class. Among other advantages, the user
provides all the parameters that may be relevant for the
classification, and  galSVM is able to purge them
and isolate only those features that are not redundant. 
The technique has been validated in different cases
\citep{hue09,hue10a}, and it has been recently
applied by \citet{hue11} to classify the full set
SDSS/DR7 galaxies with spectra at redshift $\le 0.25$, i.e., 
exactly the data base of ASK classes used 
here\footnote{
It can be downloaded from\\ {\tt http://gepicom04.obspm.fr/sdss\_morphology/Morphology\_2010.html}
}.
It requires a training set, which in this particular case 
was the morphological classification by \citet{fuk07}, 
studied separately in \S~\ref{morph_fuk}. 
\citet{hue11} defined four wide 
morphological classes that include the full range of 
possibilities: E, S0, Sab, and Scd+Im 
(see Table~\ref{equiv_morpho}). Given a galaxy, galSVM
provides the probability of belonging 
to each one of these four types, so that the galaxy will be 
regarded as E if it has a large probability of being
elliptical, and the same holds for the other types. 
The procedure classifies the $\sim 7\cdot 10^5$ galaxies of the dataset
in only a few minutes.

\begin{figure}
\centering
\includegraphics[width=0.5\textwidth]{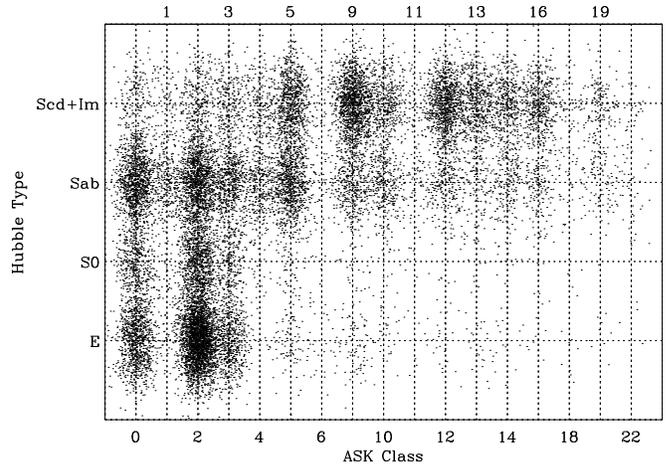}
\caption{Hubble type from \citet{hue11}  vs 
ASK class. The inference
of Hubble type uses concentration, ellipticity and color,
therefore, it is not a purely morphological classification.
However, all the trends existing in Fig.~\ref{nair2}a are
also here.
It contains only 20000 randomly selected points 
from the actual data set formed by all the $\sim 7\cdot 10^5$ 
galaxies with  redshift $< 0.25$ in SDSS/DR7.
Only major ASK classes are included.
}
\label{svm1}
\end{figure}

\citet{hue11} base the classification in three
observables, (1) the restframe colors ($g-r$ and $g-i$),
(2) the ellipticity of the galaxy images (i.e., $b/a$ with
$b$ and $a$ the minor and major axes), 
and (3) the concentration ($R_{90}/R_{50}$ in the SDSS~$i$ filter,
with  $R_\alpha$ the radius containing $\alpha$\% of the flux). 
The resulting scatter plot Hubble type vs ASK class is shown
in Fig.~\ref{svm1}. The correlation is clear and similar
to the ones described in the previous sections.
The similarity is to some extent surprising since the assigned
Hubble types are not purely morphological.  Colors
were included to derive Hubble types, and yet some morpho-types
do present a large color spread (typically, Sab galaxies).
For reasons which remain unclear to us,
the colors are regarded as unimportant by 
the galSVM algorithm (see \S~\ref{discussion} for a further analysis). 
In order to test this conjecture,
we repeated the classification in \citet{hue11},
but only with concentration and ellipticity. The
resulting scatter plot (not shown) remains similar
to Fig.~\ref{svm1}. 

On top of the overall agreement, we identify two subtle differences between  
Fig.~\ref{svm1} and Fig.~\ref{nair2}a.
First, Fig.~\ref{svm1} clearly lacks of S0 galaxies.
Somehow galSVM assigns a probability of being S0 systematically
lower than that of being E, which produces the dimming of
S0s when thresholding the sample with a constant probability
(0.5 in the case of
Fig.~\ref{svm1}, but its value is incidental in this context).
Second, the red classes ASK~1 and 4 are assigned systematically 
to spiral classes, even more sharply than in the previous plots. 
This separation results from using ellipticity
as an observable for the morphological classification.
In case of doubt, ASK~1 and 4 galaxies with isophotes of large 
eccentricity are ascribed to late morphological types. 
Recall that these two red classes seems to consist mostly of 
edge-on spirals (\S~\ref{sect_nair_abra}).

%
%
%
\subsection{Morphology from Galaxy~Zoo~1 \citep{lin10}}\label{sect_gzoo}

We have also compared the ASK classes with the Galaxy~Zoo~1 
morphological classification, where the old problem of galaxy 
classification has been addressed in a very original way using the 
current internet technology.  The morpho-types are obtained by 
public votes of more than 100000 (non-expert) volunteers  
through a specially designed web interface ({\tt http://www.galaxyzoo.org}). 
Six types are offered
to the  volunteers (E, clockwise Sp, counterclockwise Sp, edge-on, 
merger, and unknown), but this original classification is finally 
simplified to E, Sp and unknown (Un) after carrying out a significant bias 
correction to compensate for small-distant objects erroneously 
classified as E or Un.  The Galaxy Zoo team
offers a {\em clean} sample with morphologies for galaxies where, 
after debiasing, 80\% of the votes agree. 
The classification is described by \citet{lin08} whereas the 
public Galaxy Zoo release that we use is introduced in 
\citet{lin10}.

\begin{figure}
\centering
\includegraphics[width=0.5\textwidth,angle=0]{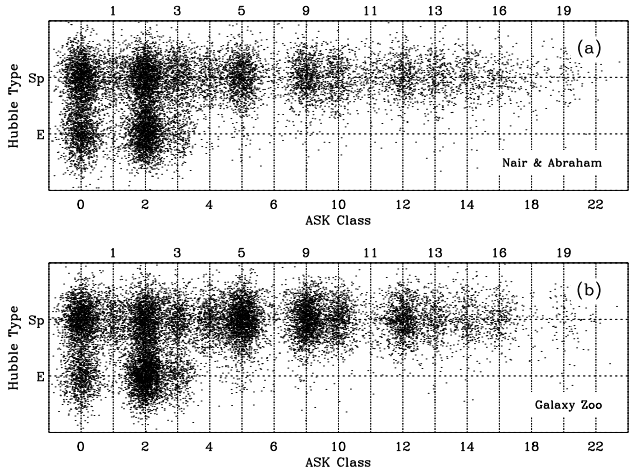}
\caption{
Comparison between the scatter plots Hubble type vs ASK class in
\citeauthor{nai10} (a) and Galaxy~Zoo~1 (b).
All spectral classes in \citet{nai10} from 
S0 to Im have been grouped into a single class Sp
to mimic the two classes considered by Galaxy Zoo. The results are 
very consistent. The number of points represented in the two plots 
is approximately the same ($\sim 15000$).
}
\label{ask_nair_zoo}
\end{figure}

The scatter plot galaxy Zoo type vs ASK class is shown
in Fig.~\ref{ask_nair_zoo}b. The classification
is coarser than the previous ones, but it is fully consistent
with them. Figure~\ref{ask_nair_zoo}a is the same as 
Fig.~\ref{nair2}a,  except that all types that are not 
ellipticals (E or ES0) have been condensed to a single class 
of spirals Sp. 
This two-class-only separation tries to mimic the galaxy Zoo 
classification.  The similarities 
between Figs.~\ref{ask_nair_zoo}a and \ref{ask_nair_zoo}b 
are more than just 
qualitative:  96.5\% of the ellipticals in 
\citet{nai10} belong to ASK~0, 2 or 3, whereas the percentage is
95.0\% for the clean sample in \citet{lin10}.
Red classes ASK~0, 2 and 3 are not exclusively 
formed by E: 49.2\% and 57\%  are spirals
in the case of Galaxy Zoo and \citet{nai10}, respectively. 
All the percentages mentioned above refer to the full match
between ASK classes and  clean Galaxy Zoo~1, which includes 247405 
galaxies.
In addition, the original (undebiased) Galaxy Zoo~1 classification
separates between Sp and edge-on systems. It proves the existence 
of red face-on spirals, 
and how ASK~1 and 4 are formed by edge-on systems. 

%
\subsection{Hubble types from the work by \citet{ken92}}\label{sec_ken}

We also considered the 53 galaxies with both integrated
spectra and morphology in the atlas by \citet{ken92}.
The results were already presented by 
\citet[][\S~7, Fig.~11]{san10} 
and will not be discussed here, except
to indicate that the relationship between morphological type and
spectroscopic class is consistent with the other
datasets considered here.

%
%

\section{ASK classes versus quantitative morphological parameters in SDSS/DR7}
\label{quantify}

\begin{figure*}
\centering
\includegraphics[width=0.7\textwidth,angle=90]{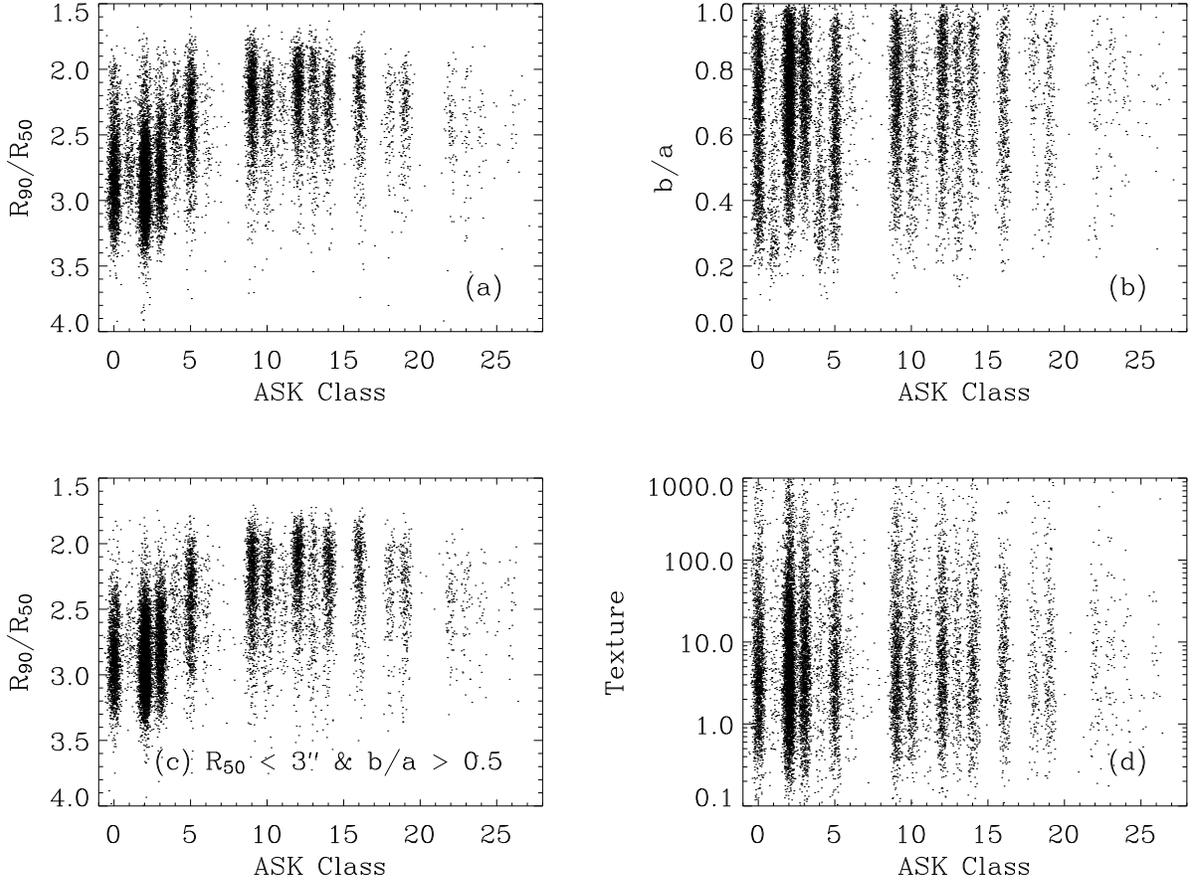}
\caption{
Scatter plots of various morphology-related parameters
vs ASK class. (a) Concentration. (Note that the ordinate 
axis has been reversed, so that that the plot looks similar 
to the plots Hubble type vs ASK class, e.g., Fig.~\ref{nair2}a.) 
(b)
Axis ratio.
(c) Concentration, but only for small ($< 3$~arcsec)
non-edge-on galaxies.
(d) Texture. 
All parameters have been computed from images in the $g$ filter.
In order to avoid overcrowding, only 10000 
representative galaxies are plotted, with their
ASK classes randomly shifted by 0.2 classes.
}
\label{ask_vs_else2a}
\end{figure*}

The photometric pipeline of SDSS provides a set of morphological 
parameters for some of the DR7 galaxies \citep[see][]{lup02,str01}.
They can be conveniently downloaded from the SDSS website, and
we have used them to explore possible relationships 
between parameters quantifying the galaxy shape, and 
the ASK classes.
We use concentration (parameterized
as $C=R_{90}/R_{50}$, i.e., the ratio between the radii 
containing 90\% and 50\% of the galaxy light), eccentricity 
(parameterized as the ratio between the minor $b$ and
major $a$ axes of the isophote at 25 mag arcsec$^{-2}$),
and {\em texture}. The interpretation
of the latter is subtle, and
corresponds to a measure of the roughness of the object, 
based on the residuals left after inverting the image and 
subtracting. We produced scatter plots of these three 
morphological parameters versus ASK class. They are shown 
in Fig.~\ref{ask_vs_else2a}.
The scatter plot $C$ versus ASK class
looks very much the same as the Hubble type versus ASK class 
(cf. Fig.~\ref{nair2}a and Fig.~\ref{ask_vs_else2a}a).
It just reflects the correlation existing between concentration
and Hubble type \citep[e.g.,][]{doi93,abr94,str01}, coupled
with the correlation between Hubble type and spectral class.
It also shows how the relationship is not one-to-one,
with many red class galaxies actually having 
a concentration characteristics of late type.
This remains valid even if only small galaxies are considered,
where the fiber used to get the galaxy spectrum covers 
a substantial fraction of the galaxy (cf. Fig.~\ref{ask_vs_else2a}a and
Fig.~\ref{ask_vs_else2a}c, the latter including only small galaxies
with $R_{50} < 3$~arcsec).
The scatter plot ellipticity vs ASK class, Fig.~\ref{ask_vs_else2a}b,  
confirms that ASK 1 and 4 are edge-on systems since $b \ll a$ in these classes. 
Moreover, there is a rather clear relationship between the 
distribution of eccentricities and the ASK\,Class.
In general, ASK~0, 2 and 3 tend to be rounded whereas
as the ASK number increases, the galaxies have 
eccentricities  more uniformly spread over the full range. 
The (log)texture represented in  Fig.~\ref{ask_vs_else2a}d
does not seem to indicate any trend with ASK class,
a lack of correlation already observed when comparing texture 
with Hubble type \citep[e.g.,][]{str01}.  
Finally, it is worth pointing out that
the scatter between quantitative morphological parameters 
and ASK classes is neither reduced not increased as compared to the 
scatter Hubble type vs ASK class 
(cf. Fig.~\ref{nair2}a and Fig.~\ref{ask_vs_else2a}a).


\section{Variation with redshift}\label{sect_redshift}

\begin{figure}
\centering
\includegraphics[width=0.5\textwidth,angle=0]{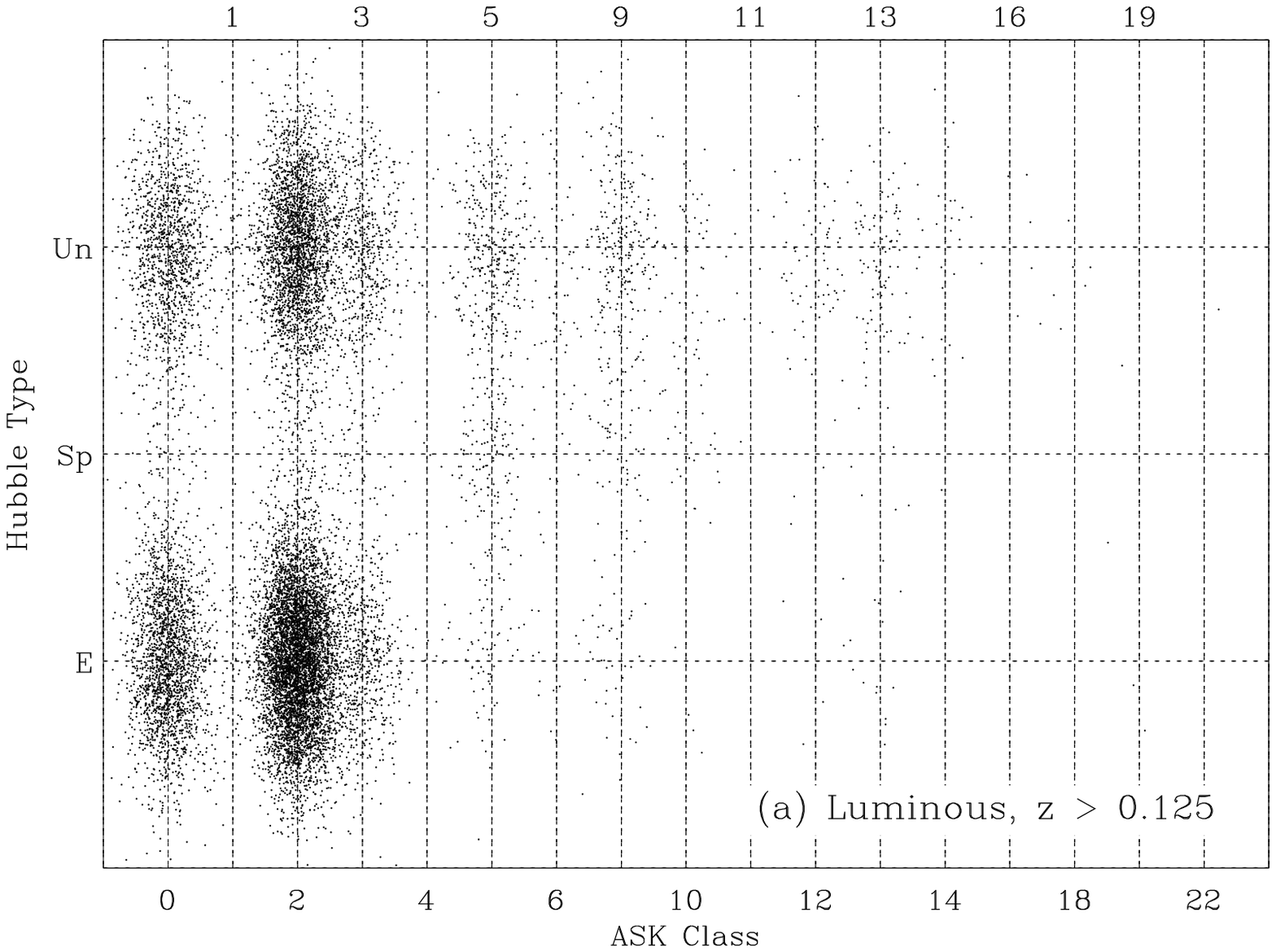}\\
\includegraphics[width=0.5\textwidth,angle=0]{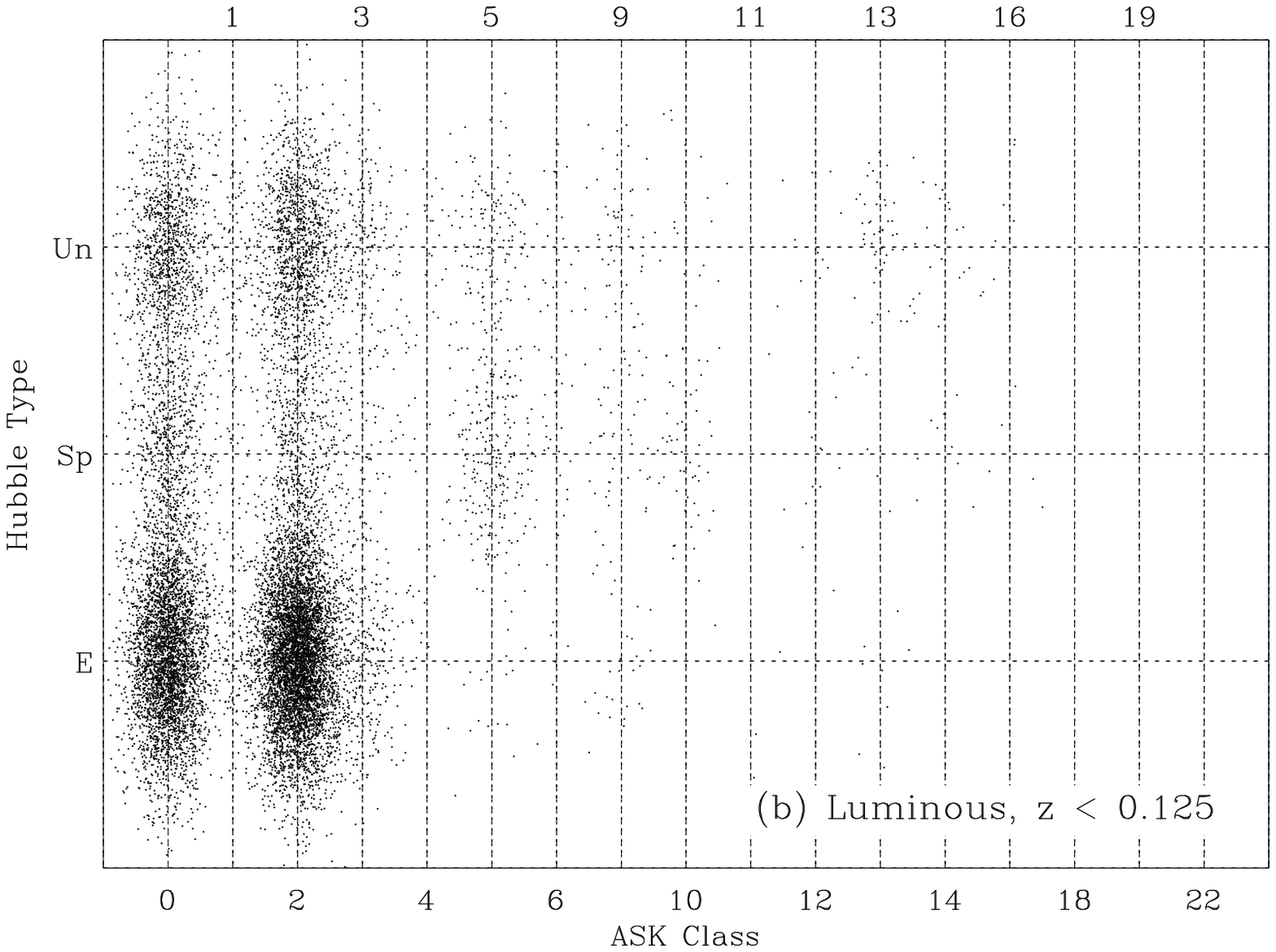}\\
\includegraphics[width=0.5\textwidth,angle=0]{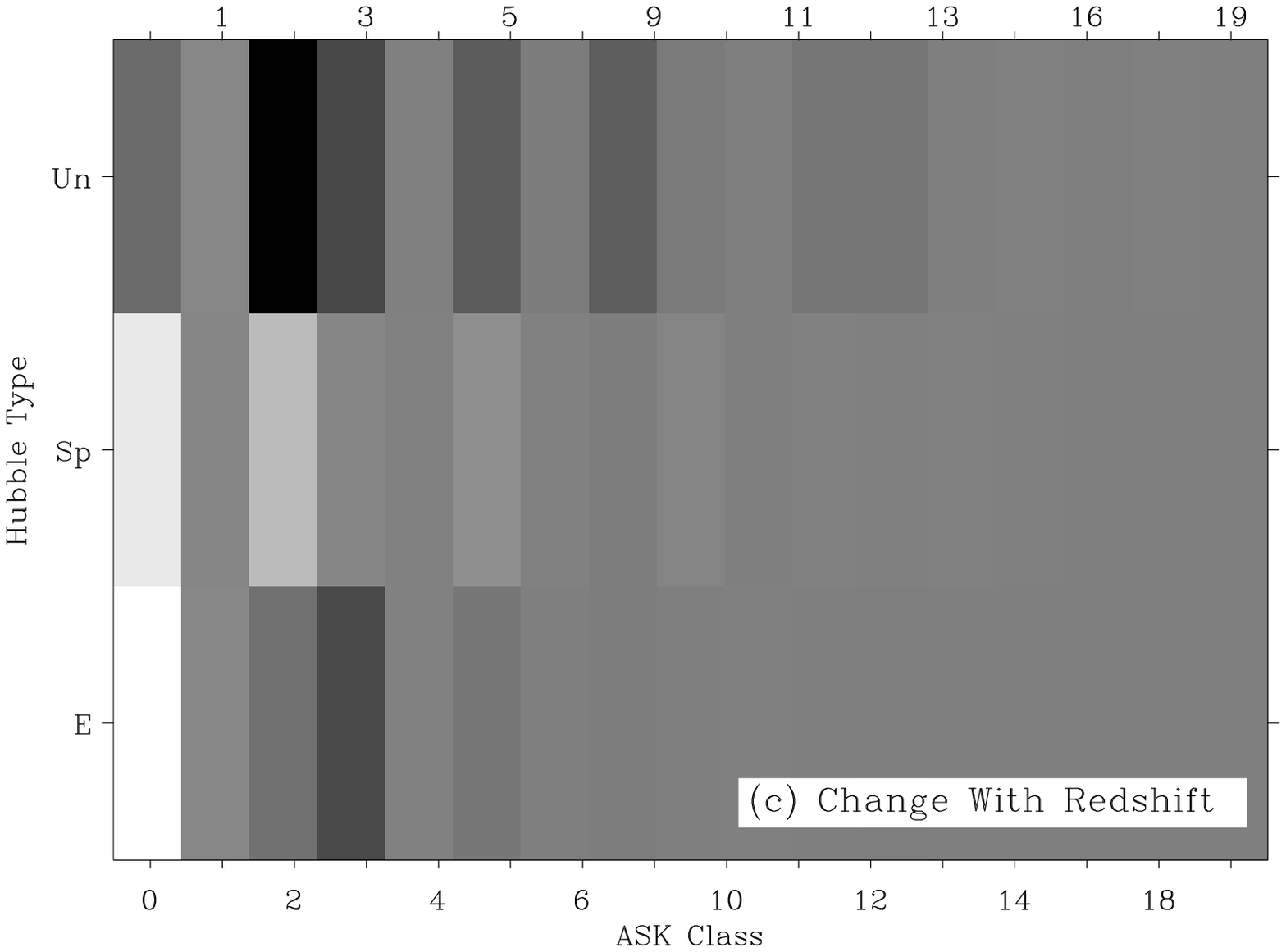}\\
\caption{
Variation with redshift of the scatter plot Hubble type 
vs ASK class. A volume limited sample of luminous 
Galaxy Zoo targets ($M_g< -22.3$) has been divided
in two redshift bins -- 
(a) from 0.125 to 0.25, and (b) from 0 to 0.125.
(c) Image with the difference between the distributions
of galaxies shown at (a) and (b) -- white corresponds to
a growth towards low redshift. 
Note the increase of Sps of all ASK classes
at low redshift, partly at the 
expense of unclassified objects (Un). However, the augment of red ASK~0 galaxies 
cannot be explained as a leakage of Un objects. We interpret it as  
the aging of the stellar populations. Panels~(a)
and (b) contain the same number of points,
and the image in (c) has been scaled between 
minimum and maximum.  
}
\label{redshift_var}
\end{figure}
The morphological classification used for reference only
includes redshifts smaller than 0.1 (\S~\ref{sect_nair_abra}). 
One needs a time baseline as long as possible to investigate changes 
with redshift, therefore, we decided to use the complete 
sample containing galaxies up to redshift $0.25$,
corresponding to
a lookback time of the order of 
3.5~Gyr\footnote{At these low redshifts the lookback time
$t$ scales linearly with redshift $z$, i.e., $t\simeq 14\,{\rm Gyr}\, z $, 
with the scaling given by the Hubble time $H_0^{-1}$.
}.
This time is long enough for the galaxies to undergo
significant changes in both the morphological 
structures and the colors --
arms do not persist longer than a few
Gyrs if the gas that feeds on-going star formation
is removed \citep[e.g.,][]{bek02}, and 
blue galaxies become red in much less than a Gyr 
if star formation shuts off \citep[e.g.,][]{bla06}.
Consequently, the temporal baseline of the analyzed SDSS/DR7
catalog is long enough to see the scatter plot 
evolving between redshift $0.25$ and the present. Do we see it? 
Figures~\ref{redshift_var}a and \ref{redshift_var}b show scatter plots
of a volume limited subsample of the Galaxy~Zoo~1 
match presented above
(\S~\ref{sect_gzoo}). We have selected all the galaxies 
bright enough to be part of the SDSS/DR7 spectroscopic
sample even if they were at redshift 0.25 (i.e., galaxies 
with  absolute $r$ magnitude smaller than $-22.3$, or
with apparent magnitude smaller than the completeness limit of SDSS even at 
redshift 0.25). The use of a volume limited sample is mandatory
since blue ASK classes are systematically
less luminous than the red ones and, therefore,  
they artificially disappear at high redshift 
from SDSS \citep[see, e.g.,][]{san10}.  
A glance at Figs. ~\ref{redshift_var}a and \ref{redshift_var}b
shows how the number of red Sps increases
in the low redshift sample at the expense of the Un. 
The change is even more evident in Fig. ~\ref{redshift_var}c, 
which displays an image with the difference between the 
histograms at low redshift and at high redshift -- 
white corresponds to an increase at low redshift. A part of 
this increase seems to be due to the improvement of angular 
resolution at low redshift. Insufficient resolution leads to small 
spirals being misclassified as Un, whereas Un tend to be classified as 
E. The final Galaxy Zoo~1 classification acknowledges
this bias, and corrects the classes accordingly \citep[][]{lin10}, 
but residuals seem to remain after correction. 
In addition to this increase of Sps, the scatter plots
in Figs. ~\ref{redshift_var}a and b
show an increase of ASK 0 galaxies at low redshift,
both of Sp and E galaxies. The increase
clearly  stands out in Fig. ~\ref{redshift_var}c.
This augment cannot be explained by the Hubble type misclassification 
described above, which would not change the galaxy ASK class.
Similarly, it cannot be ascribed to aperture effects
caused by the red centers of the galaxies
contributing more at low redshift, 
because this effect is expected to be largest
in spirals whereas the observed increase mainly affects 
E. 
We attribute the systematic reddening of the red galaxies 
as an aging of their stellar populations. 
Models for the passive evolution of large ellipticals indicate 
that their colors redden by $\Delta(g-r)\simeq - 0.2$ in 
1.5~Gyr \citep[e.g.,][]{mar09}, which is equivalent to changing 
colors from ASK~2 to ASK~0 in the time interval between the 
two redshift bins \citep[][ Table~2]{san10}. 

As far as the few blue luminous galaxies included in 
Fig.~\ref{redshift_var}, it is not at all clear whether 
they redden, or if their population
remains statistically unchanged at low redshift.
Studying the redshift variation of the blue classes requires
using the full sample, 
following an  analysis in the vein of that in appendix~\ref{appa}.
We do not address it here because the bias of the
match ASK class vs Galaxy Zoo~1 is not purely
Malmquist bias, and computing the volume associated
with each galaxy is more complex than just applying
equation~(\ref{malmquist}). The task goes beyond the scope of this 
work,  however, in view of the  variations observed
in red luminous galaxies, it is an exercise 
deserving follow up.  


%

\section{Discussion and Conclusions}\label{discussion}

There is a general trend for the local 
galaxies in red ASK classes to show early morphological
types, and for the galaxies in blue ASK classes
to have late morphological  types
(Fig.~\ref{nair1}a). However, the relationship has a large scatter with a  standard deviation 
between 2 and 3 types, both for the dispersion  of Hubble types given an ASK class, and for the 
dispersion of  spectroscopic classes fixed the Hubble type (Tables~\ref{table_ask} and 
\ref{table_hubble}, columns labeled $a$).  Figure~\ref{nair4} shows galaxy images that
illustrate the various parts of the scatter plot -- the main  trend along the diagonal, as well as the 
outliers represented by red spirals (upper left) and blue ellipticals (bottom right). 
The distributions of Hubble types given an ASK class are very skewed; they present long tails 
that go to the late morphological  types for the red galaxies, and to the  early morphological 
types for the blue spectroscopic classes.
The scatter is not produced by problems in the classification, and it is not reduced
when particular subsets are considered -- low and high galaxy masses, 
low and high density environments, barred and non-barred galaxies,
face-on galaxies, small and large galaxies, or when a volume limited
sample is considered. 

The upper left corner of the scatter plot is truly devoid of targets, i.e., 
there is a remarkable lack of Sd, Sdm,  Im and Sm  galaxies with red  spectra.
These morphological types are associated with low mass galaxies, and we
interpret this lack as an indication that, for these particular galaxies, the
time scale for the morphological changes is shorter than that 
for the  spectral changes. When these very late
type galaxies  evolve,  their Hubble type must 
become earlier before they become redder.  On the contrary, there are plenty of 
red Sa -- Sc galaxies, suggesting that a galaxy can change spectrum while
maintaining  a Hubble type, provided the galaxy is massive enough. We have 
found that 68\% of the red   galaxies (ASK~0, 2 and 3) in the catalog used for
reference \citep{nai10} are spirals rather then ellipticals. Even though red
spectra are not associated with ellipticals, most ellipticals do have red
spectra: 97\% of the ellipticals in \citet{nai10} belong to 
ASK~0, 2 or 3 whereas only   3\% of them are blue.

According to their colors, most local galaxies can be  split into red galaxies 
(red sequence) and blue galaxies \citep[blue cloud; e.g.,][]{str01,bal04,bal04b}.  
The color gap is not sharp so that in between the two extremes one finds galaxies 
with intermediate colors on the so-called green valley. These are thought to be 
transition galaxies \citep[e.g.,][]{mar07,sal07}.
  The ASK classification managed to isolate  the green~valley 
into a single  class ASK~5 \citep{san10}. We find most of ASK~5 galaxies being spiral 
(\S~\ref{sect_nair_abra}). Since the galaxies in the blue cloud are spiral as well, the green
valley galaxies  seem to be in transit from  the blue cloud to the red sequence,
with the transition  involving no major morphological change.   
The existence of red spirals mentioned above means that the  red sequence is populated by 
both spheroids  and disks.  If the galaxies in the green valley were rejuvenated  red galaxies 
with recent  star-formation activity, they should present the  mixture of E and Sp 
types characteristic of the red sequence, which is not observed. 
        %
The fact that the blue cloud preferentially contains late morphological
types but the red sequence is not dominated by early types seems to be
well established in the literature \citep[e.g.,][]{den09,bla09}.

We explore  the dependence of the scatter plot Hubble type vs 
ASK class on the environment (\S~\ref{sect_nair_abra}). \citet{nai10} provide a 
measurement of the local
density estimated from the distance to the 4th or 5th nearest neighbor, 
and we find almost no dependence of the relationship on whether low or 
high density environment galaxies are selected. Of course early types are more often present in 
dense environments, but we do not find that the relationship with spectral class 
tightens or becomes more loose in dense  environments (but see the comment on S0s below). 
This result is consistent with the idea expressed by \citet{bla09} that the environment 
affects the probability of finding early types or late types, 
but once a particular galaxy is chosen, its properties depend little on the environment.
In this sense, the  possibility of presenting varied spectra given the Hubble type 
seems to be a property of the individual galaxies, rather than being stimulated 
or directed by nearby galaxies. On top of this overall stability of the relationship, 
we note  a slight  excess of S0s with respect to Es in dense environments. 
This trend seems to agree with the finding that
S0s become relatively more frequent towards the centers of the 
clusters \citep[e.g.,][]{dre80,agu04}.

 
We investigate variation with redshift of the 
color-shape relationship using a
volume limited subsample mainly formed by luminous 
red galaxies (\S~\ref{sect_redshift}). 
From redshift $0.25$ to now some galaxies redden as expected from 
the passive  evolution of their stellar populations.

One of the ASK classes, number 6, is known to gather
active galactic nuclei \citep[AGNs,][]{san10}. 
This class seems to be formed by galaxies
with a wide range of Hubble types, from E to Sd -- 
see Fig.~\ref{nair1}a as well as Table~\ref{table_ask}, the later
showing this class to have a large negative kurtosis 
characteristic of a spreadout distribution function. 
This spread of morphological types is a known
property of AGNs \citep[e.g.][and references therein]{gab09,gri10}.

Two of the red ASK classes, 1 and 4, have been found to be dominated by 
edge-on (dust-reddened) spirals. In an analysis not mentioned in the preceding 
sections, we reproduced the template spectrum characteristic of ASK 1 and 4
with reddened spectra of bluer ASK classes. One finds a good
match of both line and continuum assuming a
one magnitude extinction at $H\beta$ distributed 
according to a Milky Way-like law
\citep{car89}.
The fact that ASK 1 and 4 are edge-on disks makes them suitable
for a number of studies -- for example, determining
of the distribution of disk thicknesses
\citep[e.g.,][]{san10b}, or studying  
physical properties of the dust associated with ongoing star 
formation.
They may also be used to calibrate weak lensing reconstructions
since we know their intrinsic (highly elongated) shapes given
the spectrum \citep[e.g.,][ and references therein]{zha10}.

The classification based on galSVM and described in 
\S~\ref{svm_sect} is, in fact, a spectro-morphological
classification. It includes colors together with other
purely morphological parameters. Surprisingly, the use of 
color does not reduce the scatter of the relationship between 
spectroscopic class and morphological type. The result is revealing,
reflecting that the scatter of colors given a Hubble type is 
intrinsic. 

The present work quantifies the relationship and scatter between 
Hubble type and spectroscopic class existing in the local universe.
It constrains the models and theories of galaxy 
evolution which should be able to account, not only for the 
global trend, but also for the scatter. In other words, the challenge 
lies in explaining how and why galaxies with the same morphology end up 
with very different spectra, and how and why galaxies with similar
spectra may have different Hubble types.

%
%
%

\begin{acknowledgements}
Thanks are due to J. Knapen and I. Garc\'\i a~de~la~Rosa
for clarifying discussions on dust-reddened edge-on Sps, and  
on bulges of large Sps misclassified as red Sps.
Thanks are also due to an anonymous referee for 
helping us improving the presentation. 
%
%
This work has been partly funded by the Spanish MICINN, project
AYA~2010-21887-C04-04.
%
JSA, ALA and CMT are members of the Consolider-Ingenio 2010 Program, grant 
MICINN CSD2006-00070: First Science with GTC.
Funding for the SDSS and SDSS-II has been provided by the Alfred P. Sloan
Foundation, the Participating Institutions, the National Science Foundation, the
U.S. Department of Energy, the National Aeronautics and Space Administration,
the Japanese Monbukagakusho, the Max Planck Society, and the Higher Education
Funding Council for England. The SDSS is managed by the Astrophysical Research 
Consortium for the Participating Institutions (for details,
see the SDSS web site at http://www.sdss.org/).
%

{\it Facilities:} \facility{Sloan (DR7, spectra)
}
\end{acknowledgements}


%
\appendix
\section{$V_{\rm max}$ approach to transform the magnitude 
limited sample of Nair \& Abraham~(2010) to  
a volume limited sample}\label{appa}

Following \citet[][\S~4]{san08},
the 2D histogram with the number of galaxies 
having ASK class $A$ and Hubble type $T$
can be written down as a sum over all the 
galaxies in the sample, 
\begin{equation}
N(A,T)=\sum_i\,\Pi\big({{A_i-A}\over{\Delta A}}\big)
\,\Pi\big({{T_i-T}\over{\Delta T}}\big).
\label{pi_eq}
\end{equation}
As usual, the symbol $\Pi$ stands for the
rectangle function,
\begin{equation}
\Pi(x)=\cases{
1&$|x|<1/2$,\cr
0&elsewhere,
}
\end{equation} 
and $\Delta T$ and $\Delta A$ represent the bin sizes of the 2D histogram.
Rather than the histogram of observed galaxies, $N(A,T)$, 
we want the histogram to be obtained if galaxies were drawn from
a volume limited sample, ${\Large n}(A,T)$. It can be
estimated as 
\begin{equation}
n(A,T)=\sum_i\,\Pi\big({{A_i-A}\over{\Delta A}}\big)
\,\Pi\big({{T_i-T}\over{\Delta T}}\big){1\over{V_{i\,{\rm max}}}}
\label{vmax}
\end{equation}
where $V_{i\,{\rm max}}$ is the maximum volume at which the galaxy $i$
could have been detected in our observation. $V_{i\,{\rm max}}$ depends
on properties of the galaxy $i$, as well as on the biases affecting
the observation. Equation~(\ref{vmax}) implicitly assumes the volume 
sampled by the  observation to be uniform, with the observational biases
to be such that they allow us to detect all types of galaxies.
We compute the maximum volume as
\begin{equation}
V_{i\,{\rm max}}={W\over3}10^{-{3\over{5}}(M_i+9)}\,{\rm Mpc}^3,
\label{malmquist}
\end{equation}
with $M_i$ the absolute magnitude of the galaxy, and $W$ the solid angle 
covered by the survey.
The previous expression assumes the observation to be 
magnitude limited (which is approximately correct for the sample, 
with a limit magnitude of 16), 
and neglects the bias associated with the surface brightness threshold 
(see the references given in \citeauthor{san08}~\citeyear{san08}).

\end{document}